\documentclass[runningheads]{llncs}
\usepackage{eccv}
\usepackage{eccvabbrv}
\usepackage{graphicx}
\usepackage{booktabs}
\usepackage{enumitem}
\usepackage[misc]{ifsym}
\newcommand{\modelname}{ProCT\xspace}
\newcommand{\modelnamedd}{ProCT$^\dagger$\xspace}

\newcommand{\papertitle}{Prompted Contextual Transformer for Incomplete-View CT Reconstruction}
\newcommand{\abbrpapertitle}{ProCT for Incomplete-View CT Reconstruction}

\newcommand{\vct}[1]{\boldsymbol{#1}} 
\newcommand{\mat}[1]{\boldsymbol{#1}} 

\renewcommand{\paragraph}[1]{\noindent\textbf{#1}\quad}
\newcommand{\qryname}{\mathrm{src}}
\newcommand{\conname}{\mathrm{con}}

\newcommand{\ndet}{{N_\mathrm{det}}}
\newcommand{\nfview}{{N^\mathrm{full}_\mathrm{view}}}
\newcommand{\nview}{{N_\mathrm{view}}}

\newcommand{\conv}{f_\mathrm{conv}}
\newcommand{\linear}{f_\mathrm{lin}}
\newcommand{\mlp}{f_\mathrm{mlp}}

\newcommand{\relu}{r}

\newcommand{\sinox}{\mat{S}_\text{in}}
\newcommand{\sinoy}{\mat{S}}
\newcommand{\imgx}{\mat{X}}
\newcommand{\imgy}{\mat{Y}}
\newcommand{\imgp}{\widehat{\imgy}}

\newcommand{\qryin}{\imgx}
\newcommand{\qrypred}{\imgp}
\newcommand{\qryout}{\imgy}
\newcommand{\coninx}{\widetilde{\imgx}}
\newcommand{\coniny}{\widetilde{\imgy}}

\newcommand{\mixqryin}{\mat{F}}
\newcommand{\mixconin}{\widetilde{\mat{F}}}

\newcommand{\mixqrykey}{\mat{K}}
\newcommand{\mixqryquery}{\mat{Q}}
\newcommand{\mixqryvalue}{\mat{V}}
\newcommand{\mixconvalue}{\widetilde{\mat{V}}}

\newcommand{\mixspat}{\mat{F}^\mathrm{spat}}
\newcommand{\mixfreq}{\mat{F}^\mathrm{freq}}

\newcommand{\mixqrysf}{\mat{G}_\mathrm{intr}}
\newcommand{\mixconout}{\widetilde{\mat{G}}}

\newcommand{\mixqryout}{\mat{G}}

\newcommand{\qrycond}{\vct{p}}
\newcommand{\concond}{\widetilde{\vct{p}}}

\usepackage[pagebackref,breaklinks,colorlinks,citecolor=eccvblue]{hyperref}
\usepackage{orcidlink}

\begin{document}
\title{\papertitle} 
\titlerunning{\abbrpapertitle}

\author{Chenglong Ma\inst{1} \and
Zilong Li\inst{1} \and
Junjun He\inst{2} \and
Junping Zhang\inst{1} \and
Yi Zhang\inst{3}\textsuperscript{(\Letter)} \and
Hongming Shan\inst{1}\textsuperscript{(\Letter)}
}

\institute{Fudan University \and Shanghai AI Laboratory \and Sichuan University\\
\email{clma22@m.fudan.edu.cn} \quad \email{yzhang@scu.edu.cn} \quad \email{hmshan@fudan.edu.cn}}

\maketitle

\begin{abstract}
Incomplete-view computed tomography (CT) can shorten the data acquisition time and allow scanning of large objects, including sparse-view and limited-angle scenarios, each with various settings, such as different view numbers or angular ranges. 
However, the reconstructed images present severe, varying artifacts due to different missing projection data patterns. 
Existing methods tackle these scenarios/settings \emph{separately} and \emph{individually}, which are cumbersome and lack the flexibility to adapt to new settings. 
To enjoy the multi-setting synergy in a single model, we propose a novel \underline{Pro}mpted \underline{C}ontextual \underline{T}ransformer (\modelname) for incomplete-view CT reconstruction. 
The novelties of \modelname lie in two folds.
First, we devise a projection view-aware prompting to provide setting-discriminative information, enabling a single \modelname to handle diverse incomplete-view CT settings. 
Second, we propose artifact-aware contextual learning to sense artifact pattern knowledge from in-context image pairs, making \modelname capable of accurately removing the complex, unseen artifacts. 
Extensive experimental results on two publicly available clinical CT datasets  demonstrate  the superior  performance of \modelname over state-of-the-art methods---including single-setting models---on a wide range of incomplete-view CT settings, strong transferability to  unseen datasets and scenarios, and improved performance when sinogram data is available.
The source code will be made publicly available.

\keywords{Incomplete-view CT \and Prompt learning \and In-context learning}
\end{abstract}

\begin{figure}[tb]
  \centering
  \includegraphics[width=\linewidth]{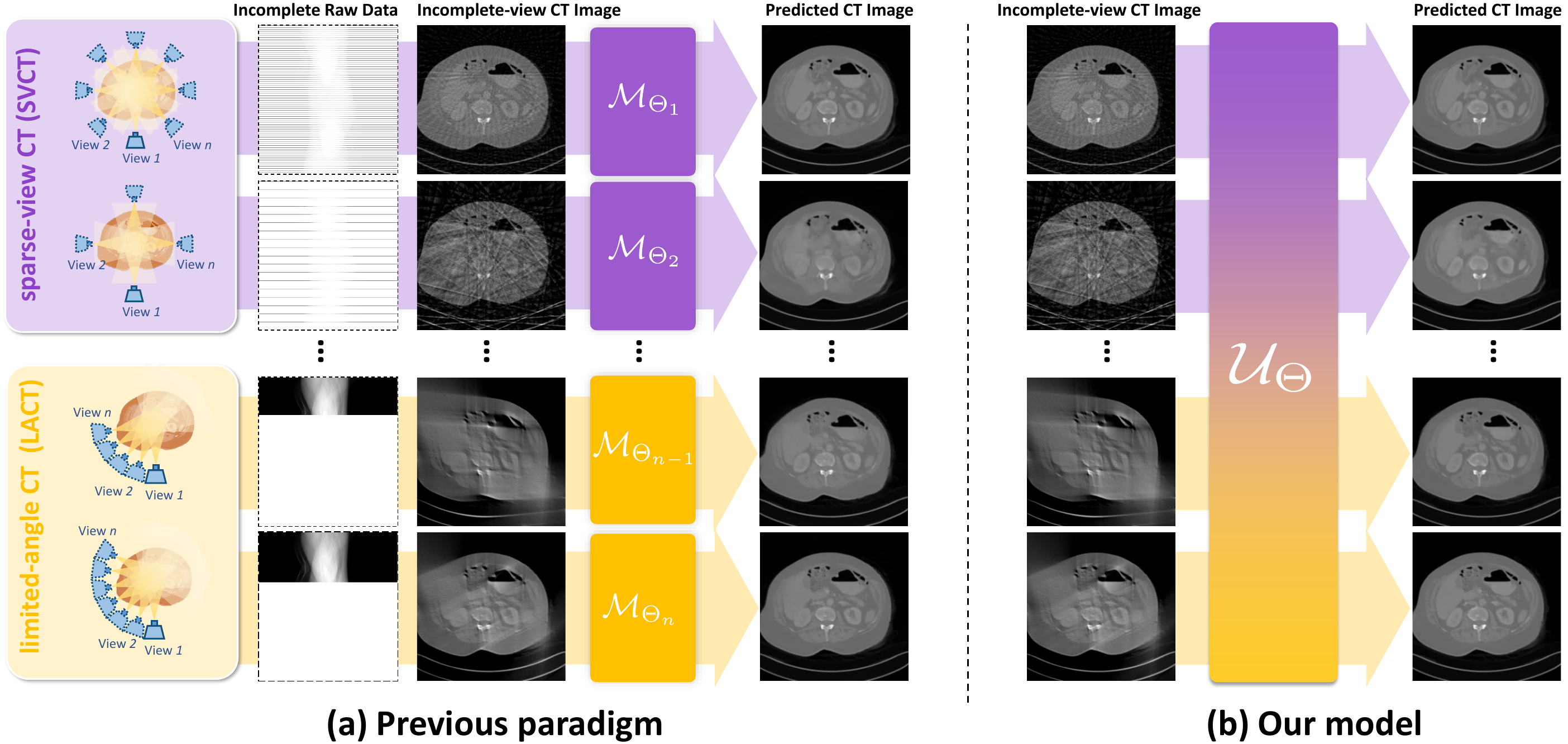}
  \caption{The proposed \modelname enables all-in-one incomplete-view CT reconstruction with a single model, exploiting multi-setting synergy to achieve state-of-the-art performance while saving computational resources.}
  \label{fig:teaser}
\end{figure}

\section{Introduction}\label{sec:intro}
X-ray computed tomography (CT) is a cornerstone in the field of medical imaging, providing valuable insights into the internal structures of human body non-invasively. This technique acquires x-ray raw data (\aka projection data, or sinogram) from different views around the subject, which are then used to reconstruct cross-sectional images~\cite{ct-outlook}. Despite the benefits, conventional CT imaging acquires a full set of raw data from various views, leading to prolonged scanning time and increased radiation dose. In addition, some patients with physical limitations make a full-view scan impractical.  

Incomplete-view CT has emerged as a promising solution to accelerating scanning time~\cite{cho2013motion}, reducing radiation dose, and accommodating physical constraints or irregular scenarios in medical imaging~\cite{arai2001dental,sheng2020sequential}, which only acquires a small set of projection data. This strategy can be further categorized as sparse-view CT (SVCT) and limited-angle CT (LACT), depending on different missing data (or sampling) patterns shown in \cref{fig:teaser}. Concretely, for fan-beam CT, typical SVCT settings sample projection from equidistant views covering $[0^\circ, 360^\circ]$, while LACT selects projections from views that cover a specific range of angles. Both SVCT and LACT result in a fairly small number of views compared to the predetermined full-view number. 
In either case, there exists a range of settings tailored to specific clinical requisites, and the incomplete raw data can lead to distinct artifacts globally distributed on the image reconstructed by conventional filtered back-projection (FBP) algorithms~\cite{wang2013meaning}. 

Existing methods handle these settings in isolation, requiring the training of a separate model for each setting. However, this ``one-model-for-one-setting'' fashion, while straightforward, yields models lacking the capacity to harness the synergy among multiple settings. This consequently results in the following issues that hinder their broader application. 
\emph{First}, single-setting models exhibit a noticeable deficit in robustness when confronted with even subtle variations in the CT setting they are trained for. This is due to the potentially amplified overfitting to one single incomplete-view sampling pattern, which is particularly evident for dual-domain models that combine knowledge from both sinogram- and image-domain data.
\emph{Second}, single-setting models often struggle to adapt to unseen settings during practical use. Although possible, pre-configuring multiple single-setting models does not fundamentally solve the problem, which also introduces unnecessary training efforts and storage overhead and leaves gaps for new, uncovered settings.

Alternatively, one can train a model with data from as many settings as possible. However, this is nontrivial due to the mutual interference of drastically different artifacts~\cite{gloredi}. 
Recent advances in large language models~\cite{gpt3,prompt-survey,icl-survey} offer the possibility of utilizing prompting method and in-context data to manipulate the model's parameter space during multi-setting training and to achieve flexible adaptation on new settings. 

In this paper, we present \underline{Pro}mpted \underline{C}ontextual \underline{T}ransformer (\modelname), a robust model to perform all-in-one CT reconstruction across diverse incomplete-view CT settings, as depicted in \cref{fig:teaser}. 
The novelties of our \modelname are two-fold. \emph{First}, to empower \modelname with the reconstruction ability on a wide range of fine-grained incomplete-view settings, we propose the view-aware prompting technique to inject the setting-discriminative information into the network. \emph{Second}, to capture the complex artifact pattern within each setting, we propose artifact-aware contextual learning that leverages the artifact guidance from another incomplete- and full-view CT image pair to remove the artifact in the input incomplete-view CT image. 
In doing so, \modelname can enjoy the multi-setting synergy, achieving robust and transferable incomplete-view CT reconstruction, outperforming the state-of-the-art methods. We also show that \modelname can be further enhanced once the sinogram data are available.

In summary, our contributions are listed as follows.
\begin{itemize}[topsep=0pt]
    \item[1)] We present the prompted contextual transformer or \modelname to solve the challenging task of incomplete-view CT reconstruction with diverse settings, using a single model in one pass. 
    \item[2)] We propose view-aware prompting to provide setting-discriminative knowledge, which endows \modelname with the capability to robustly handle a wide range of incomplete-view CT images within a single model. 
    \item[3)] We propose artifact-aware contextual learning to sense the artifact pattern information from an incomplete- and full-view CT phantom pair, which can ensure a better grasp of the artifacts within each setting. 
    \item[4)] Extensive experimental results demonstrate the robustness and superiority of \modelname over state-of-the-art reconstruction methods across diverse incomplete-view CT scenarios and settings. Remarkably, our \modelname is a \emph{single} model to handle all settings compared with existing methods that require different models for different settings.
\end{itemize}

\section{Related Work}\label{sec:related}
\paragraph{Learning-based incomplete-view CT reconstruction.}
Existing learning-based methods for incomplete-view CT reconstruction mainly include image-domain methods and dual-domain ones. Image-domain methods~\cite{fbpconvnet,ddnet,gloredi} take incomplete-view CT images as input and treat the reconstruction as an image post-processing task. On the other hand, dual-domain methods utilizes both the incomplete-view sinogram data and image data. 

Generally, dual-domain methods excel in accurately restoring incomplete-view CT images compared to simple image-domain methods, as they leverage information from two domains. Some approaches~\cite{dreamnet,learn++,ayad2024} unroll the conventional iterative algorithms into deep learning frameworks, others~\cite{dudotrans,cross,freeseed} perform sinogram inpainting and image post-processing simultaneously. 

Unfortunately, these methods are designed in a single-setting manner and lack transferability and flexibility. They can be sensitive to tiny perturbation in the predetermined setting and require re-training once the clinical practice opts to a new CT scenario or setting. 
Few prior studies looked into this problem~\cite{shu2022sparse,ayad2022tomographic}. For instance, Shu \etal~\cite{shu2022sparse} proposed using a deep image prior~\cite{dip} to tackle SVCT and LACT with one model. However, their method is computationally intensive, needing hundreds of iterations to process every single image. Unlike the existing work, our model can leverage synergy among various incomplete-view CT settings and achieve universal and robust incomplete-view CT reconstruction with fine-grained settings in one pass.

\paragraph{Universal models for image restoration.}Inspired by the remarkable generalizability of large language models and generation models, recent research has turned to resolving various tasks~\cite{transweather} in the field of image restoration. 
Notably, prompt-based methods and in-context learning have shown their great potential. 
Prompt-based methods aim to condition the inputs with task-related knowledge without altering the model parameters, offering great flexibility. Pioneering works such as PromptIR~\cite{promptir} and ProRes~\cite{prores} utilize prompts for universal image restoration and demonstrate their effectiveness in low-level vision. 
In-context learning~\cite{gpt3}, on the other hand, aims to teach the model to imitate the given example and improve the generalizability in a few-shot fashion. For example, MAE-VQGAN~\cite{mae-vqgan} and Painter~\cite{painter} can reason on the query input guided by contextual input-output examples, thereby performing various visual tasks using a single model. 

In contrast, we propose incorporating the \emph{view-aware} prompting technique and \emph{artifact-aware} contextual learning for all-in-one incomplete-view CT reconstruction, where the view prompt encodes the semantic of the CT sampling view distribution, and contextual learning provides the complementary image-domain information for better grasping the complex artifact patterns. 

\section{Method}\label{sec:method}
\subsection{Problem Formulation and motivation}
Typical CT scanning produces a series of x-ray sinogram data (\aka raw data) as an object is rotated within a scanner. The full-view sinogram,  $\sinoy \in\mathbb{R}^{\nfview \times \ndet}$, represents the attenuation of x-rays through the object from various views, where $\nfview$ denotes the number of full views and $\ndet$ the number of detectors. 
Through back-projection algorithm $\mathcal{B}$ (\eg, FBP), $\sinoy$ can be used to reconstruct a CT image $\imgy=\mathcal{B}(\mat{S})$ visualizing the inner structures of the object.

As shown in~\cref{fig:teaser}, in the incomplete-view CT scenarios, we only use a small portion of the full-view sinogram $\sinoy$ for imaging. Concretely, we use a view sampling vector $\vct{v}\in\{0,1\}^{\nfview}$ to indicate which projection views are sampled from $\sinoy$; \ie, the $1$ values in $\vct{v}$ indicate the view distribution. Obviously, the $1$ values in $\vct{v}$  are equidistantly distributed for SVCT and localized on a contiguous segment for LACT.
With $\vct{v}$, we can create the incomplete-view sinogram, $\sinox=\mathcal{R}(\mathrm{diag}(\vct{v})\cdot\sinoy)$, where $\mathcal{R}$ is a reduction operator to remove the zero rows, and $\sinox \in\mathbb{R}^{\nview \times \ndet}$ has much fewer valid views (\ie, $\nview \ll \nfview$). 

The filtered back-projection is then performed to produce the incomplete-view CT image $\imgx =\mathcal{B}(\sinox)$. 
Unfortunately, this becomes a severe ill-posed problem when $\nview \ll \nfview$~\cite{louis1989incomplete,frikel2013lact}, and the resulting image $\imgx$ presents severe and distinct artifacts given different sampling vector $\vct{v}$. The goal of incomplete-view CT reconstruction is to produce an image $\imgp$ from $\imgx$ (or $\sinox$ if the sinogram is available) with a level of quality close to the ground-truth full-view image $\imgy$. 

Existing methods typically address the clinical settings in both SVCT and LACT separately due to their distinct artifact patterns resulting from different sampling view distribution, which lacks the robustness on seen settings as well as the transferability to unseen settings and creates an unnecessary drain on computing resources. Therefore, we are motivated to build a \emph{single} model to achieve robust and transferable incomplete-view CT reconstruction.

\subsection{Overview of \modelname}

\begin{figure*}[t] 
    \centering \includegraphics[width=1.0\textwidth]{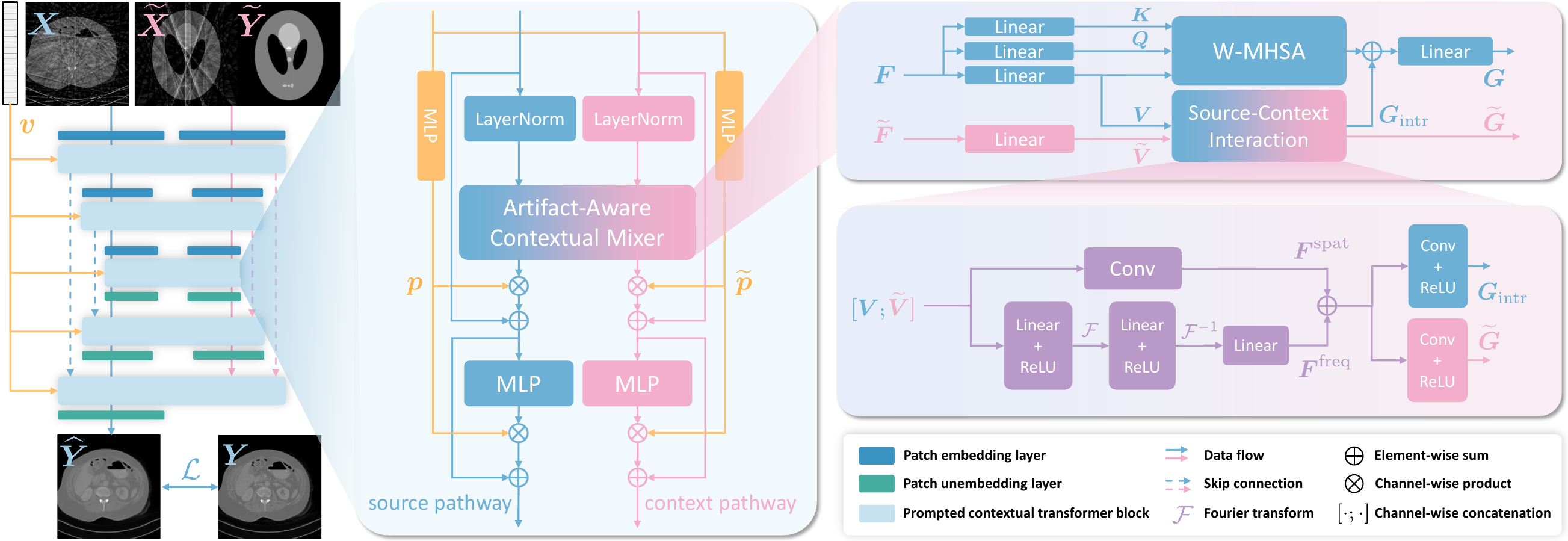}
    \caption{Overview of the proposed \modelname. Detailed configurations can be found in Sec.~\textcolor{red}{A.1} of our supplementary material.}
    \label{fig:model} 
\end{figure*}

\cref{fig:model} depicts the overall hourglass architecture of our proposed model, \modelname, consisting of several prompted contextual transformer blocks with a source pathway and a context pathway. \modelname incorporates two key designs: \emph{view-aware prompting} and \emph{artifact-aware contextual learning}.

View-aware prompting is crucial to achieving multi-setting incomplete-view CT reconstruction, using a view prompt $\mat{p}$ to inject the setting-discriminative information into the model. However, $\mat{p}$ only indicates the location of the artifacts, and it is still difficult to learn the complex artifact patterns. 
Therefore, we further propose artifact-aware contextual learning to leverage the complementary information of artifact patterns from an incomplete- and full-view in-context pair $\mathcal{C}$. The reconstruction process of our \modelname $\mathcal{U}_\Theta$ can be formulated by
\begin{equation}
    \qrypred = \mathcal{U}_\Theta(\qryin, \mathcal{C}, \vct{p}),
\end{equation}
where $\qryin$ denotes the input source incomplete-view CT image to be reconstructed, and $\mathcal{C}=[\coninx;\coniny]$ denotes the given in-context pair. Here, $\coninx$ and $\coniny$ are the incomplete- and full-view CT images, respectively, and $[\cdot;\cdot]$ represents the concatenation along the channel dimension.

We highlight that the view prompt can either be derived from the view sampling vector or be parameterized by some learnable embeddings, and the in-context pair can be created from the commonly used CT phantom or other publicly available incomplete-view CT image datasets. We choose the CT phantom as our default setting for its easy availability. In the following, we detail these two key designs.

\subsection{View-aware Prompting}\label{ssec:prompter}
Prompting techniques have been applied to universal image restoration as a powerful indicator of various restoration tasks~\cite{promptir,prores}, where 
a learnable prompt is dedicated to one task interacting with the input images or latent features. Nonetheless, in the realm of incomplete-view CT, we are dealing with a vast range of fine-grained settings (\eg, the angular range in LACT can be any reasonable sub-interval within $[0^\circ, 360^\circ]$), making it impractical to assign a separate prompt for each setting. Furthermore, these incomplete-view CT settings exhibit intrinsic relations (\eg, the sampling patterns of two different settings can overlap). Treating them independently may fail to capture such relations and the resulting artifact locations in the image. 

To encode such fine-grained CT settings and their intrinsic relations, we turn to the sampling vector $\vct{v}$ characterizing the difference between these settings to construct our view prompt.
$\vct{v}$ is initially used to create the incomplete-view sinograms; some examples are illustrated in~\cref{fig:teaser}. 

To improve expressivity without introducing excessive complexity, we simply apply two individual prompters to create view prompts $\qrycond = \mlp^\qryname(\vct{v})$ and $\concond = \mlp^\conname(\vct{v})$ for the source and context pathways, respectively. For simplicity, we implement the prompters as multi-layer perceptrons (MLPs).
These prompts modulate the feature outputs from both the mixer and MLP within each transformer block. For instance, let $\mat{H}\in\mathbb{R}^{C\times L}$ denote the $C$-channel input for source pathway MLP, the MLP output $\mat{H}_\mathrm{mlp}\in\mathbb{R}^{C\times L}$ is modulated by $\qrycond\in\mathbb{R}^C$:
\begin{align}
    \mat{H}_\mathrm{mlp} &= \mat{H} + \qrycond \otimes \mlp^\qryname(\mat{H}),
\end{align}
where $\otimes$ is the broadcasting channel-wise product. Modulations elsewhere are similar, as shown in~\cref{fig:model}. We note that $\vct{v}$ can also be parameterized as a learnable vector with simple modification, and we leave this in Sec.~\textcolor{red}{C.1} of our supplementary material.

\subsection{Artifact-aware Contextual Learning}\label{ssec:token-mixer}
The view prompt encodes the semantics of the view distribution and indicates the location of the artifacts. However, learning complex artifact patterns in a wide range of incomplete-view CT settings remains challenging. 

To address this issue, we seek help from an incomplete- and full-view CT \emph{phantom} pair that can provide complementary information on the typical artifact pattern given a specific CT setting without relying on the sinogram data. Specifically, we design an artifact-aware contextual mixer to effectively combine informative spatial and frequency features from the contextual pair with features extracted from the source image. As depicted in~\cref{fig:model}, the contextual mixer is comprised of a windowed multi-head self-attention (W-MHSA)~\cite{swin,dehazeformer} branch and an interaction branch. 

\paragraph{W-MHSA branch.}This branch extracts the essential features from the input source image itself. Given the source feature map $\mixqryin$, we first generate the key ($\mixqrykey$), query ($\mixqryquery$) and value ($\mixqryvalue$), and then compute the W-MHSA features, which will later be fused with the features from the interaction branch. 

\paragraph{Interaction branch.}This branch performs spatial-frequency interaction between the input source feature map $\mixqryin$ and the contextual feature map $\mixconin$, considering that utilizing both spatial and frequency information greatly benefits CT image reconstruction~\cite{gloredi,freeseed}. As shown in~\cref{fig:model},
we first linearly project $\mixconin$ as  $\mixconvalue$ and then concatenate it with $\mixqryvalue$ as the input to  two parallel interaction blocks.

One block applies convolution over $\mixqryvalue$ and $\mixconvalue$ to produce the spatial interaction feature map $\mixspat$. Another block performs convolution in the Fourier domain to obtain frequency interaction feature map $\mixfreq$. Finally, we obtain the interaction source feature map $\mixqrysf$ and  corresponding contextual feature map $\mixconout$ using two individual convolution layers:
\begin{align}
    \mixqrysf = \relu\big(
    \conv^\qryname(
    \mixspat + \mixfreq
    )\big), \quad
    \mixconout = \relu\big(
    \conv^\conname(
    \mixspat + \mixfreq
    )\big),
\end{align}
where $\relu(\cdot)$ represents the ReLU activation. 

\paragraph{Fusing source-interaction information.} We further fuse interaction feature map $\mixqrysf$ and the source W-MHSA feature map as follows:
\begin{equation}
    \mixqryout = \linear\big(
    \sigma\big(
    \mat{Q}\mat{K}^\top / \sqrt{d}
    \big)\mat{V} + \mixqrysf \big),
\end{equation}
where $d$ is the query/key dimension, $\sigma(\cdot)$ denotes the softmax function, and $\linear$ is the linear layer. For brevity,  reshaping and windowing operations are omitted. The output source feature map $\mixqryout$ and  contextual feature map $\mixconout$ are then passed through MLPs, serving as the input to subsequent transformer blocks.

\subsection{\modelnamedd: A Flexible Dual-Domain Extension}
Benefiting from its image-domain design, \modelname can serve as a frozen pre-trained model and be seamlessly integrated with prevalent dual-domain learning frameworks. In this paper, we extend \modelname to its dual-domain counterpart, denoted by \modelnamedd. Different from the conventional dual-domain methods that apply a cascade design, \modelnamedd utilizes a parallel architecture with its image-domain sub-network (\ie, \modelname) and sinogram-domain sub-network. Information from the two domains is then synergized by a fusion module to reconstruct the final results. To keep \modelnamedd simple, we adopt the SE-Net in DuDoNet~\cite{dudonet} as our sinogram completion model and a lightweight U-Net~\cite{unet} as the fusion module. Please refer to Sec.~\textcolor{red}{A.2} for the network and training details.

Compared with the cascaded dual-domain models, this parallel architecture enjoys greater flexibility and adaptability, facilitating the incorporation of prior knowledge in either domain.

\subsection{Training and Inference of \modelname}\label{ssec:loss}
For training, to remove the artifacts while preserving the global structures in the output CT image, we use the combination of  $\ell_1$ loss $\mathcal{L}_\mathrm{\ell1}$ and multi-scale structural similarity loss~\cite{ms-ssim} $\mathcal{L}_\mathrm{ssim}$, defined as:
\begin{equation}
    \mathcal{L}(\qrypred,\qryout) = \mathcal{L}_\mathrm{\ell1}(\qrypred,\qryout) + \alpha\mathcal{L}_\mathrm{ssim}(\qrypred,\qryout),
    \label{eq:single-loss}
\end{equation}
where $\alpha\geq0$ is the balancing factor, empirically set to 0.1. We train our model \modelname $\mathcal{U}_\Theta$ by minimizing the following loss function over the incomplete-view CT scenarios $T$ and settings $t$:
\begin{align}
    \underset{\Theta}{\min}
     \sum_{T\in\mathcal{T}}
    \mathbb{E}_{t\sim\mathcal{A}_T}
    \mathbb{E}_{(\qryin, \qryout, \mathcal{C}, \mat{p})\in\mathcal{D}_t} \left[s_t\mathcal{L}(\qrypred, \qryout)
    \right],\label{eq:optim}
\end{align}
where $s_t$ is an introduced scaling factor to trade-off learning difficulties among different settings, and $\mathcal{T}$ is an index set for different scenarios. Please refer to Sec.~\textcolor{red}{C.2} for discussion of the loss scaling schedule $s_t$. In this paper, $\mathcal{T}=\{1,2\}$, with $T=1$ for SVCT and $T=2$ for LACT. $\mathcal{A}_T$ is a collection of settings under the scenario indexed by $T$. We have $\mathcal{A}_1$ representing a set of different numbers of views $\nview$ for SVCT, and $\mathcal{A}_2$ representing a set of different angular ranges $R$ for LACT. These settings are uniformly selected in each iteration, and each of them corresponds to a dataset $\mathcal{D}_t$ with different incomplete-view CT artifacts.

For inference, the input source image, an in-context image pair, and a view prompt are fed into the \modelname. We note that the in-context pair is constructed from the readily available CT phantom or public CT image datasets, and that the view prompt can be easily obtained from the CT scanning protocol without involving any commercial privacy issues. Hence, \modelname does not require additional information when tested on a new unseen dataset.

\section{Experiments}\label{sec:experiments}

\subsection{Experimental setup}\label{ssec:exp-setup}

\begin{table*}[!t]
\centering
\caption{Quantitative evaluation [PSNR (db), SSIM (\%)] for state-of-the-art methods on the DeepLesion dataset. The best results are highlighted in \textbf{bold} and the second best results are \underline{underlined}. ``\textbf{AVG.}'': average  values for one CT scenario.}
\resizebox{1.0\textwidth}{!}{
\begin{tabular}{lcccccccccccccccccc}
\toprule
{} & \multicolumn{10}{c}{\textbf{SVCT}} & \multicolumn{8}{c}{\textbf{LACT}} \\
\cmidrule(lr){2-11} 
\cmidrule(lr){12-19}
{} & \multicolumn{2}{c}{$\nview=18$} & \multicolumn{2}{c}{$\nview=36$} & \multicolumn{2}{c}{$\nview=72$} & \multicolumn{2}{c}{$\nview=144$} & \multicolumn{2}{c}{\textbf{AVG.}} & \multicolumn{2}{c}{$R=[0^\circ,90^\circ]$} & \multicolumn{2}{c}{$R=[0^\circ,120^\circ]$} & \multicolumn{2}{c}{$R=[0^\circ,150^\circ]$} & \multicolumn{2}{c}{\textbf{AVG.}} \\
\cmidrule(lr){2-3}
\cmidrule(lr){4-5}
\cmidrule(lr){6-7}
\cmidrule(lr){8-9}
\cmidrule(lr){10-11}  
\cmidrule(lr){12-13}
\cmidrule(lr){14-15}
\cmidrule(lr){16-17}
\cmidrule(lr){18-19}  
{\textbf{Method}} & PSNR & SSIM & PSNR & SSIM & PSNR & SSIM & PSNR & SSIM & PSNR & SSIM & PSNR & SSIM & PSNR & SSIM & PSNR & SSIM & PSNR & SSIM \\
\midrule
\multicolumn{19}{c}{single-setting models} \\
\midrule
FBP 
&21.07 &32.01 &24.49 &44.00 &29.58 &60.52 &35.16 &79.76 &27.58 &54.07 &18.92 &52.96 &22.63 &64.12 &27.45 &74.22 &23.00 &63.77 \\
DDNet~\cite{ddnet} 
&33.46 &85.51 &35.10 &89.87 &40.06 &94.82 &43.36 &97.03 &38.00 &91.81 &32.67 &89.60 &37.32 &93.24 &41.38 &96.00 &37.12 &92.95 \\
FBPConvNet~\cite{fbpconvnet} 
&34.17 &87.79 &36.99 &90.85 &40.48 &94.43 &44.01 &97.66 &38.91 &92.68 &34.86 &91.87 &38.93 &93.95 &43.16 &96.40 &38.98 &94.07 \\
DuDoTrans~\cite{dudotrans} 
&34.22 &88.01 &37.12 &92.74 &40.94 &95.91 &\underline{44.74} &\underline{97.90} &39.25 &93.64 &33.05 &90.83 &37.95 &94.74 &42.94 &\textbf{97.65} &37.98 &94.59 \\
CROSS~\cite{cross} 
&34.27 &87.99 &37.40 &92.44 &41.29 &96.09 &44.20 &97.57 &39.26 &93.52 &35.35 &91.61 &39.07 &95.04 &43.24 &97.03 &39.22 &94.56 \\
FreeSeed\textsubscript{\textsc{dudo}}~\cite{freeseed} 
&34.46 &88.72 &37.67 &92.90 &\underline{41.36} &\underline{96.31} &\textbf{44.80} &\textbf{97.92} &39.62 &93.98 &34.47 &90.80 &39.12 &94.42 &43.70 &97.12 &39.10 &94.11 \\
GloReDi~\cite{gloredi} 
&\underline{35.20} &88.45 &\textbf{38.50} &\textbf{94.19} &40.92 &96.01 &44.21 &97.53 &39.71 &\underline{94.00} &36.29 &91.67 &39.33 &93.31 &43.60 &96.67 &39.74 &93.88 \\
\midrule
\multicolumn{19}{c}{multi-setting models} \\
\midrule
FreeSeed$^*_\textsc{dudo}$
&26.71 &58.36 &30.26 &70.60 &34.64 &80.81 &35.03 &88.36 &31.66 &74.53 &27.97 &78.46 &33.06 &86.08 &34.89 &90.27 &31.97 &84.94 \\
GloReDi$^*$
&25.52 &54.83 &29.07 &67.92 &31.87 &77.09 &32.77 &83.53 &29.81 &70.94 &27.38 &81.27 &35.07 &88.01 &36.78 &90.82 &33.08 &86.70 \\
UniverSeg~\cite{universeg}
&31.69 &82.54 &35.18 &88.19 &39.32 &93.06 &42.34 &96.39 &37.13 &90.05 &32.15 &87.45 &35.15 &87.45 &35.85 &91.23 &38.79 &92.82 \\
\modelname (ours) &\textbf{35.48} &\underline{89.00} &38.20 &92.98 &41.14 &96.10 &44.11 &97.63 &\underline{39.73} &93.82 &\underline{37.83} &\underline{92.64} &\textbf{41.46} &\underline{95.15} &\textbf{44.70} &96.99 &\underline{41.33} &\underline{94.93} \\
\modelnamedd (ours) &35.05 &\textbf{91.24} &\underline{38.34} &\underline{93.71} &\textbf{41.48} &\textbf{96.59} &44.29 &97.84 &\textbf{39.78} &\textbf{94.85} &\textbf{38.69} &\textbf{93.02} &\underline{41.44} &\textbf{95.43} &\underline{43.97} &\underline{97.29} &\textbf{41.36} &\textbf{95.25} \\
\bottomrule
\end{tabular}
}
\label{tab:comp-deeplesion}
\end{table*}

\paragraph{Datasets.}We conduct our experiments on two publicly available clinical CT image datasets:  DeepLesion dataset~\cite{deeplesion}, and  AAPM dataset~\cite{aapmmyo}. 
For DeepLesion dataset, we randomly choose 10,000 images from 3,012 patients as the training set and another 1,000 images from 24 patients as the test set. For AAPM dataset, 1,145 images from 2 patients are selected as the external test set. Images in both datasets are split based on patients without information leakage between training and testing, and have a resolution of 256 $\times$ 256.

\paragraph{Data preparation.}The TorchRadon toolbox~\cite{torch-radon} is employed to simulate the fan-beam CT routine with $\nfview=720$. For SVCT, we select $\nview$ as in $\mathcal{A}_1=[9,288]$ for training, and the test settings are from $\{18,36,72,144\}$. For LACT, the collection of angular ranges for training is $\mathcal{A}_2=\{[0^\circ,r]: r\in[60^\circ,180^\circ]\}$, while the test settings are from $\{[0^\circ,90^\circ],[0^\circ,120^\circ],[0^\circ,150^\circ]\}$. To exemplify, $\nview=72$ in SVCT means that 72 views are equidistantly sampled from full 720 views spanning $[0^\circ, 360^\circ]$ to create the incomplete-view CT images, while an angular range $R=[0^\circ,90^\circ]$ only samples sinogram from the views of $[0^\circ,90^\circ]$ out of $[0^\circ, 360^\circ]$. Please see Sec. \textcolor{red}{B.1} for more details.

\paragraph{Competing methods.}We compare our \modelname and \modelnamedd with the following methods in various incomplete-view CT scenarios and settings: DDNet~\cite{ddnet}, FBPConvNet~\cite{fbpconvnet}, DuDoTrans~\cite{dudotrans}, CROSS~\cite{cross}, FreeSeed\textsubscript{\textsc{dudo}}~\cite{freeseed}, and GloReDi~\cite{gloredi}. DDNet, FBPConvNet, and GloReDi are powerful image-domain methods for incomplete-view CT, while DuDoTrans, CROSS, and FreeSeed\textsubscript{\textsc{dudo}} are state-of-the-art dual-domain methods. UniverSeg~\cite{universeg} is a newly proposed model for universal medical image segmentation, but we find it also applicable to our task and train it using the same strategy as our \modelname. By default, we follow the corresponding literature to set the hyperparameters for each method. Specifically, to adapt GloReDi to the LACT scenario, the ``intermediate angular range'' for knowledge distillation during its training is set to $[0^\circ, a+30^\circ]$ if the input LACT image corresponds to an angular range of $[0^\circ, a]$. Note that one would have to train multiple versions of these competing methods for different settings, but our models only need to be trained \emph{once} to cover them all.

\paragraph{Implementation detail.}Our models are implemented using PyTorch~\cite{pytorch} framework. We train \modelname for 70 epochs using an Adam optimizer~\cite{adam} with $(\beta_1,\beta_2)=(0.5, 0.999)$. For \modelname, we only need to train one single model by solving the problem in Eq.~\eqref{eq:optim}. Note that $\mathcal{A}_1$ and $\mathcal{A}_2$ cover a vast range of fine-grained incomplete-view CT settings. To enhance the stability in the training process, we train \modelname on a few settings with the proposed loss scaling schedule for the first 40 epochs, and then move on to all settings in $\mathcal{A}_1$ and $\mathcal{A}_2$, following an ``easier-to-harder'' strategy. The learning rate starts from $10^{-4}$ and is halved for every 20 epochs. For \modelnamedd, we keep the image-domain sub-network (\ie, \modelname) frozen and train the rest components with another 40 epochs. For other competing methods, we individually train one copy for each setting $t\in\mathcal{A}_T$ on an NVIDIA RTX 3090 GPU for same epochs with a mini-batch size of 2. 

\begin{figure*}[t]
    \centering
    \includegraphics[width=1\linewidth]{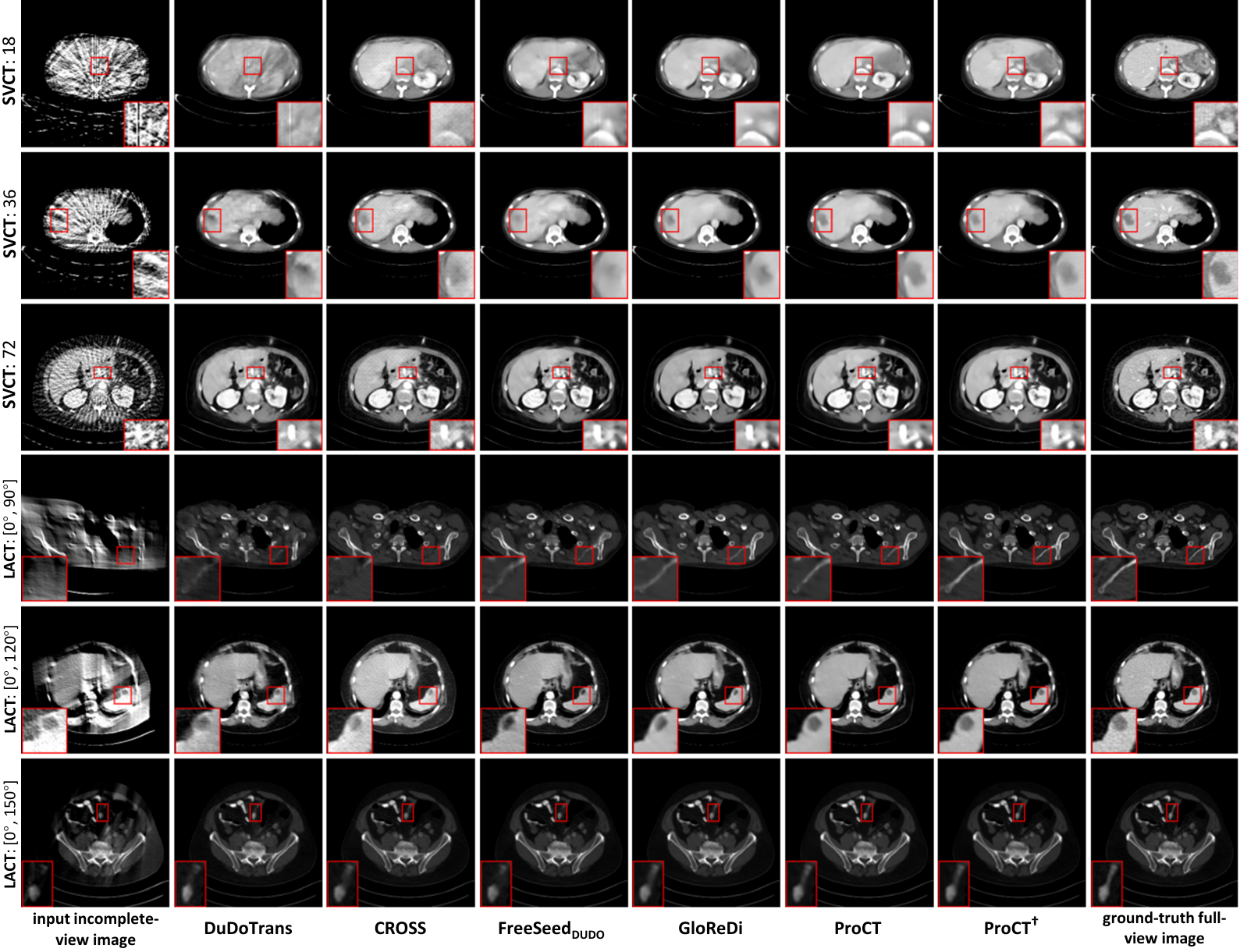}
    \caption{Visual comparison of state-of-the-art methods (from left to right) on DeepLesion dataset over different incomplete-view scenarios and settings (from top to bottom). The display windows are set to [-115, 185] HU for Rows 1, 2, 3 and 5, and [-150, 850] HU for Rows 4 and 6. Regions of interest are zoomed in for better viewing.}
    \label{fig:comp-deeplesion}
\end{figure*}

\paragraph{Evaluation metrics.}We employ peak signal-to-noise ratio (PSNR) for pixel-wise accuracy evaluation and structural similarity (SSIM)~\cite{iqa} for perceived visual quality evaluation. For these two metrics, the higher, the better.

\subsection{Comparison with state-of-the-art models}
We compare our \modelname and \modelnamedd with state-of-the-art single-setting models in the SVCT and LACT scenarios on the DeepLesion test set, as shown in~\cref{tab:comp-deeplesion}.

\paragraph{Quantitative results.}\cref{tab:comp-deeplesion} shows that, in a setting where relatively more views are given (\eg, $\nview=144$ for SVCT; an angular range of $R=[0^\circ, 150^\circ]$ for LACT), dual-domain methods exhibit clear superiority over image-domain ones due to the rich information provided by the sinogram. However, this superiority would diminish with fewer views. We also find that the LACT scenario is more challenging, where sinogram completion poses a more difficult extrapolation problem rather than a simpler interpolation one, which is the typical case for the SVCT scenario. This is more evident for dual-domain methods: note that an angular range $R=[0^\circ, 90^\circ]$ in LACT indicates a 87.5\% drop of sinogram-domain information, but dual-domain models like DuDoTrans and FreeSeed\textsubscript{\textsc{dudo}} perform even worse than their corresponding SVCT versions in $\nview=18$ that indicates a 97.5\% drop. This is because the unsuccessful sinogram extrapolation in those dual-domain methods without particular designs may accumulate errors, leading to limited performance gains. 

In contrast, our models outperform the single-setting models in most of the settings. Once the raw data is available, \modelname can be flexibly extended to \modelnamedd to utilize the sinogram information, achieving the best performance in terms of average metrics. Interestingly, \modelnamedd demonstrates superior SSIM performance compared to \modelname by leveraging data from two domains.

We also assess whether the ability of all-in-one incomplete-view CT reconstruction can be simply unlocked by applying our multi-setting training strategy. For the most performing single-setting models FreeSeed\textsubscript{\textsc{dudo}} and GloReDi, we develop two variants, namely GloReDi$^*$ and FreeSeed$^{*}_{\textsc{dudo}}$, both of which are trained using the same multi-setting strategy as with our models. GloReDi$^*$ and FreeSeed$^{*}_{\textsc{dudo}}$ have the same architectures as their single-setting counterparts. As shown in Rows 8 and 9 of~\cref{tab:comp-deeplesion}, FreeSeed$^{*}_{\textsc{dudo}}$ and GloReDi$^*$ perform worse than their single-setting counterparts, indicating that the non-trivial transferability of \modelname stems from within our model rather than the training strategy. This observation is also in line with previous findings~\cite{gloredi} that, without proper design, training with incomplete-view CT data under different CT settings can adversely impact the learning process and lead to poor performance. In contrast, with the view prompt and the artifact-aware contextual learning, our models are able to leverage the multi-setting synergy and achieve better performance in most of the incomplete-view CT settings.

\paragraph{Qualitative results.}\cref{fig:comp-deeplesion} shows the reconstructed incomplete-view CT images by different methods, where the regions of interest are zoomed in to aid visualization. One can observe that, while previous dual-domain methods achieve higher quantitative results in certain settings, they can fail to recover the detail in the degraded CT images or produce distorted results, such as the aorta in Row 1, the bone structure in Row 4, and the splenic lesion in Row 5. More visualization examples can be found in our supplementary material.

\subsection{Evaluation on unseen dataset and scenario}
\paragraph{Generalizing on a new dataset.}We apply the above models trained on the DeepLesion dataset to the AAPM test set without any fine-tuning for further evaluation of the generalizability. The results are shown in~\cref{tab:comp-aapm}. While dual-domain models like FreeSeed$_{\textsc{dudo}}$ typically generalize worse on the challenging LACT scenario than on the SVCT scenario, the proposed \modelname achieves superior average performance for both scenarios, showing robust generalizability to a new dataset. \modelnamedd exhibits a slight performance gap compared to \modelname, possibly due to potential overfitting to the sinogram domain of the DeepLesion dataset. This susceptibility to sinogram domain becomes clear when a subtle variation occurs in the predicted sinogram and introduces stubborn secondary artifacts. Despite this, \modelnamedd still achieves competitive reconstruction results.

\begin{table*}[!t]
\centering
\caption{Quantitative evaluation [PSNR (db), SSIM (\%)] for the generalizability of all methods on the AAPM dataset. The best results are highlighted in \textbf{bold} and the second best results are \underline{underlined}. ``\textbf{AVG.}'': average values for one CT scenario.}
\resizebox{1.0\textwidth}{!}{
\begin{tabular}{lcccccccccccccccccc}  
\toprule
{} & \multicolumn{10}{c}{\textbf{SVCT}} & \multicolumn{8}{c}{\textbf{LACT}} \\
\cmidrule(lr){2-11} 
\cmidrule(lr){12-19}
{} & \multicolumn{2}{c}{$\nview=18$} & \multicolumn{2}{c}{$\nview=36$} & \multicolumn{2}{c}{$\nview=72$} & \multicolumn{2}{c}{$\nview=144$} & \multicolumn{2}{c}{\textbf{AVG.}} & \multicolumn{2}{c}{$R=[0^\circ,90^\circ]$} & \multicolumn{2}{c}{$R=[0^\circ,120^\circ]$} & \multicolumn{2}{c}{$R=[0^\circ,150^\circ]$} & \multicolumn{2}{c}{\textbf{AVG.}} \\
\cmidrule(lr){2-3}
\cmidrule(lr){4-5}
\cmidrule(lr){6-7}
\cmidrule(lr){8-9}
\cmidrule(lr){10-11}  
\cmidrule(lr){12-13}
\cmidrule(lr){14-15}
\cmidrule(lr){16-17}
\cmidrule(lr){18-19}  
{\textbf{Method}} & PSNR & SSIM & PSNR & SSIM & PSNR & SSIM & PSNR & SSIM & PSNR & SSIM & PSNR & SSIM & PSNR & SSIM & PSNR & SSIM & PSNR & SSIM \\
\midrule
\multicolumn{19}{c}{single-setting models} \\
\midrule
FBP 
&22.46 &35.68 &25.71 &48.56 &30.89 &66.52 &36.33 &84.50 &28.85 &58.82 &18.78 &58.89 &23.44 &70.17 &28.85 &78.54 &23.69 &69.20 \\
DDNet~\cite{ddnet} 
&34.74 &90.70 &37.61 &92.17 &41.39 &95.99 &44.42 &97.49 &39.54 &94.09 &33.20 &92.04 &37.37 &95.57 &41.67 &97.34 &37.55 &94.98 \\
FBPConvNet~\cite{fbpconvnet} 
&35.00 &89.70 &38.06 &92.58 &41.75 &95.54 &45.18 &97.95 &40.00 &93.94 &33.60 &92.60 &38.95 &96.19 &43.14 &97.52 &38.56 &95.43 \\
DuDoTrans~\cite{dudotrans} 
&35.29 &91.47 &38.61 &94.15 &42.20 &96.68 &\textbf{45.86} &\textbf{98.22} &40.49 &95.13 &31.36 &91.07 &36.91 &95.82 &42.31 &98.13 &36.86 &95.00 \\
CROSS~\cite{cross} 
&35.00 &91.32 &37.92 &93.84 &39.82 &95.80 &43.05 &97.46 &38.94 &94.61 &34.89 &93.54 &38.49 &95.72 &43.67 &97.77 &39.01 &95.67 \\
FreeSeed\textsubscript{\textsc{dudo}}~\cite{freeseed} 
&35.35 &91.51 &38.61 &94.35 &\underline{42.23} &96.73 &\underline{45.77} &\underline{98.19} &40.48 &95.20 &32.65 &92.85 &38.37 &\underline{96.74} &43.68 &\textbf{98.27} &38.23 &95.95 \\
GloReDi~\cite{gloredi} 
&\underline{36.06} &92.18 &\underline{39.47} &\textbf{95.16} &41.85 &96.62 &44.71 &97.64 &40.52 &95.40 &35.13 &94.26 &39.18 &96.57 &43.81 &97.95 &39.38 &96.26 \\
\midrule
\multicolumn{19}{c}{multi-setting models} \\
\midrule
UniverSeg~\cite{universeg} 
&33.09 &89.53 &36.83 &92.81 &41.01 &95.85 &44.21 &97.58 &38.78 &94.94 &31.61 &91.60 &36.15 &95.35 &40.13 &97.18 &35.96 &94.71 \\
\modelname (ours) &\textbf{36.38} &\textbf{92.76} &\textbf{39.48} &\underline{95.01} &\textbf{42.35} &\textbf{96.78} &45.34 &97.96 &\textbf{40.89} &\textbf{95.63} &\textbf{36.74} &\textbf{95.04} &\textbf{40.50} &\textbf{97.03} &\textbf{44.29} &\underline{98.05} &\textbf{40.51} &\textbf{96.70} \\
\modelnamedd (ours) &35.65 &\underline{92.36} &39.10 &94.78 &\textbf{42.35} &\underline{96.76} &45.05 &97.90 &\underline{40.54} &\underline{95.45} &\underline{35.90} &\underline{94.75} &\underline{40.48} &\underline{97.00} &\underline{44.06} &98.01 &\underline{40.15} &\underline{96.59} \\
\bottomrule
\end{tabular}
}
\label{tab:comp-aapm}
\end{table*}

\begin{figure}[t] 
    \centering \includegraphics[width=1.0\linewidth]{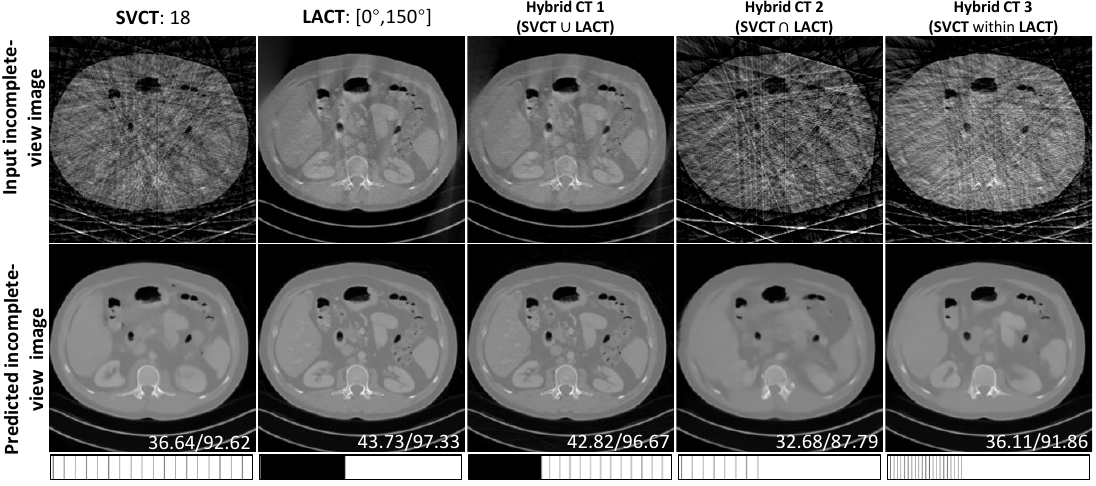}
    \caption{\modelname can easily adapt to out-of-domain hybrid CT scenarios. PSNR/SSIM values are displayed in the corner of the image; The last row shows the view prompt.}
    \label{fig:hybrid-visualization} 
\end{figure}

\paragraph{Adapting to out-of-domain scenarios.}Further, we find that \modelname can also generalize on or adapt to new hybrid CT scenarios. Here, we manipulate an LACT angular range of $[0^\circ, 150^\circ]$ and an SVCT number of views of $18$ to create three out-of-domain hybrid CT scenarios: ``Hybrid CT 1'' which is the union of LACT and SVCT, ``Hybrid CT 2'' which is the intersection, and ``Hybrid CT 3'' which performs SVCT within the angular range of LACT. We evaluate the performance of \modelname on these scenarios.

The incomplete-view images in the ``Hybrid CT 1'' scenario have only subtle differences compared with the ones in the $[0^\circ, 150^\circ]$-LACT scenario, so we directly apply our well-trained \modelname to it without re-training. However, the other two scenarios exhibiting pronounced artifacts are much more difficult to generalize. To solve this problem, we only fine-tune the prompters for 3 epochs on the DeepLesion dataset to create two versions specifically for ``Hybrid CT 2'' and ``Hybrid CT 3'', while all other parameters of \modelname are frozen. A typical CT sample is shown in~\cref{fig:hybrid-visualization}. Despite the complex artifacts in the input hybrid CT image, \modelname still manages to remove the artifact and restore the image content. Note that adapting \modelname to all these unseen scenarios requires minimal or no fine-tuning.

\subsection{Ablation studies}
We validate the effectiveness of our designs by comparing the default version of \modelname and other variants in~\cref{tab:ab-all}. The configurations are as follows: (1) default version (``phantom pair w/ $\mat{p}$''); (2) removing view prompt from the default (``default w/o $\mat{p}$''); (3) replacing phantom pair in the default version with random real incomplete- and full-view CT image pairs in the DeepLesion training set (``real pair w/ $\mat{p}$''); (4) replacing phantom pair with the input incomplete-view CT image, \ie, no extra information is provided (``no context w/ $\mat{p}$''); (5) replacing phantom pair with an incomplete-view phantom image (``incomplete-view phantom w/ $\mat{p}$''); and (6) replacing phantom pair with a full-view phantom image (``full-view w/ $\mat{p}$''). 

\paragraph{Ablation on the view prompt.}\label{ssec:exp-prompt}By comparing Rows 1 and 2 of~\cref{tab:ab-all}, we observe that the variant without the view prompt is still aware of different incomplete-view CT settings, thanks to the artifact pattern information in the in-context pair. Unfortunately, this information is limited for effectively discriminating the fine-grained settings, leading to sub-optimal performance. In contrast, the proposed view prompt benefits our model by injecting the setting-discriminative information. With proper modification, we can also set the view prompts to be learnable embeddings during training, which is discussed in Sec. \textcolor{red}{C.1} of our supplementary material.

\begin{table}
    \centering
    \caption{Ablation on different configurations.}
    \resizebox{\linewidth}{!}{
    \begin{tabular}{lccccc}
    \toprule
        Configurations
        & SVCT-PSNR & SVCT-SSIM 
        & LACT-PSNR & LACT-SSIM \\
    \midrule
    (1) phantom pair w/ $\vct{p}$ (default) & \underline{39.73} & \underline{93.82} & \textbf{41.33} & \underline{94.93} \\
    (2) phantom pair w/o $\vct{p}$ & 37.69 & 91.68 & 37.79 & 92.47 \\
    (3) real pair w/ $\vct{p}$ & \textbf{39.80} & \textbf{93.90} & \underline{41.32} & \textbf{95.07} \\
    (4) no context w/ $\vct{p}$ & 39.57 & 93.23 & 41.00 & 94.49 \\
    (5) incomplete-view phantom w/ $\vct{p}$ & 38.27 & 90.55 & 39.29 & 92.62 \\
    (6) full-view phantom w/ $\vct{p}$ & 38.07 & 91.20 & 39.44 & 93.34 \\
    \bottomrule
    \end{tabular}}
    \label{tab:ab-all}
\end{table}

\paragraph{Ablation on the contextual learning.}\label{ssec:exp-context}Rows 1 and 4 of~\cref{tab:ab-all} demonstrates the effectiveness of artifact-aware contextual learning. While CT phantom is chosen as our default context considering its easy access, the in-context pair can also be created from real CT image pairs, when they are available.

Row 3 in~\cref{tab:ab-all} shows that choosing real CT image pairs from the DeepLesion dataset as context leads to better performance. We attribute this improvement to the increased artifact similarity between the in-context pairs and the source CT images, as the image pair is derived from real CT images and may offer a more accurate guidance for the model to adeptly mitigate the signal-dependent artifacts in the input incomplete-view CT images. 
By comparing  Rows 1 and 6 in~\cref{tab:ab-all}, we also find that either the single incomplete-view phantom or the full-view one alone performs even worse than using no contextual information, which indicates that the effectiveness of our model may be attributed to the utilization of incomplete- and full-view pairs as context that aids the learning of mapping from degraded to clean images. 

Despite the benefit of real CT image pairs, we opt for CT phantom as our in-context pair since it is more reliably accessible in clinical practice, offering a pragmatic solution when real patient data is unavailable.

\section{Conclusion and Discussion}\label{sec:conclusion}
In this paper, we present \modelname, an all-in-one artifact removal model to tackle the challenging task of incomplete-view CT reconstruction with diverse practical settings. By incorporating the view-aware prompting and artifact-aware contextual learning, \modelname can utilize multi-setting synergy to resolve a wide range of incomplete-view CT reconstruction tasks using a single image-domain model, with robustness on the variation of the seen setting and transferability to complex unseen artifacts. Extensive experimental results show that \modelname does not require the raw data, yet achieves competitive or even better reconstruction performance compared to the dominating dual-domain methods. When the raw data are accessible, \modelname can be flexibly extended to a dual-domain model to leverage the extra information for further enhancements. 

We acknowledge some limitations in this work. First, the dual-domain extension model, \modelnamedd, exhibits reduced generalizability on unseen datasets, due to the challenging out-of-domain sinogram completion problem. Future improvements may involve advanced dual-domain learning designs, \eg, cascaded designs with parameter-efficient fine-tuning techniques like LoRA~\cite{lora}. Second, the view-aware prompting can be extended beyond the sampling vector (\eg, anatomies, dose levels, other physical parameters, \etc) to encode richer information for more robust reconstruction. Third, rather than digital phantom, using physical phantom or real-world data to construct the in-context pair can provide better guidance for artifact removal, as they share more similar artifact patterns with the input image. In the future, we will look into these issues and build a more robust and accurate model for incomplete-view CT reconstruction.

\newpage
\bibliographystyle{splncs04}

\begin{thebibliography}{10}
\providecommand{\url}[1]{\texttt{#1}}
\providecommand{\urlprefix}{URL }
\providecommand{\doi}[1]{https://doi.org/#1}

\bibitem{arai2001dental}
Arai, Y., Honda, K., Iwai, K., Shinoda, K.: Practical model {3DX} of limited
  cone-beam x-ray {CT} for dental use. In: International Congress Series.
  vol.~1230, pp. 713--718 (2001)

\bibitem{ayad2024}
Ayad, I., Nicolas, L., Nguyen, M.K.: Qn-mixer: A quasi-newton mlp-mixer model
  for sparse-view {CT} reconstruction. In: Proceedings of the IEEE/CVF
  Conference on Computer Vision and Pattern Recognition (2024)

\bibitem{ayad2022tomographic}
Ayad, I., Tarpau, C., Cebeiro, J., Nguyen, M.K.: Tomographic reconstruction
  from sparse-view and limited-angle data using a generative adversarial
  network. In: 2022 16th International Conference on Signal-Image Technology \&
  Internet-Based Systems. pp. 341--347. IEEE (2022)

\bibitem{mae-vqgan}
Bar, A., Gandelsman, Y., Darrell, T., Globerson, A., Efros, A.A.: Visual
  prompting via image inpainting. In: Advances in Neural Information Processing
  Systems (2022)

\bibitem{gpt3}
Brown, T., Mann, B., Ryder, N., Subbiah, M., Kaplan, J.D., Dhariwal, P.,
  Neelakantan, A., Shyam, P., Sastry, G., Askell, A., Agarwal, S.,
  Herbert-Voss, A., Krueger, G., Henighan, T., Child, R., Ramesh, A., Ziegler,
  D., Wu, J., Winter, C., Hesse, C., Chen, M., Sigler, E., Litwin, M., Gray,
  S., Chess, B., Clark, J., Berner, C., McCandlish, S., Radford, A., Sutskever,
  I., Amodei, D.: Language models are few-shot learners. In: Advances in Neural
  Information Processing Systems. vol.~33, pp. 1877--1901 (2020)

\bibitem{universeg}
Butoi, V.I., Ortiz, J.J.G., Ma, T., Sabuncu, M.R., Guttag, J., Dalca, A.V.:
  Univer{S}eg: Universal medical image segmentation. International Conference
  on Computer Vision  (2023)

\bibitem{cho2013motion}
Cho, J.H., Fessler, J.A.: Motion-compensated image reconstruction for cardiac
  {CT} with sinogram-based motion estimation. In: 2013 IEEE Nuclear Science
  Symposium and Medical Imaging Conference. pp.~1--5 (2013)

\bibitem{icl-survey}
Dong, Q., Li, L., Dai, D., Zheng, C., Wu, Z., Chang, B., Sun, X., Xu, J., Sui,
  Z.: A survey for in-context learning. arXiv preprint arXiv:2301.00234  (2022)

\bibitem{frikel2013lact}
Frikel, J., Quinto, E.T.: Characterization and reduction of artifacts in
  limited angle tomography. Inverse Problems  \textbf{29}(12),  125007 (2013)

\bibitem{cross}
Hu, D., Zhang, Y., Quan, G., Xiang, J., Coatrieux, G., Luo, S., Coatrieux,
  J.L., Ji, X., Han, H., Chen, Y.: {CROSS}: Cross-domain
  residual-optimization-based structure strengthening reconstruction for
  limited-angle {CT}. IEEE Transactions on Radiation and Plasma Medical
  Sciences  \textbf{7}(5),  521--531 (2023)

\bibitem{lora}
Hu, E.J., Shen, Y., Wallis, P., Allen-Zhu, Z., Li, Y., Wang, S., Wang, L.,
  Chen, W.: Lo{RA}: Low-rank adaptation of large language models. In:
  International Conference on Learning Representations (2022)

\bibitem{fbpconvnet}
Jin, K.H., McCann, M.T., Froustey, E., Unser, M.: Deep convolutional neural
  network for inverse problems in imaging. IEEE Transactions on Image
  Processing  \textbf{26}(9),  4509--4522 (2017)

\bibitem{adam}
Kingma, D.P., Ba, J.: {Adam}: A method for stochastic optimization. arXiv
  preprint arXiv:1412.6980  (2014)

\bibitem{noise2noise}
Lehtinen, J., Munkberg, J., Hasselgren, J., Laine, S., Karras, T., Aittala, M.,
  Aila, T.: {N}oise2{N}oise: Learning image restoration without clean data. In:
  Proceedings of the 35th International Conference on Machine Learning.
  Proceedings of Machine Learning Research, vol.~80, pp. 2965--2974 (10--15 Jul
  2018)

\bibitem{gloredi}
Li, Z., Ma, C., Chen, J., Zhang, J., Shan, H.: Learning to distill global
  representation for sparse-view {CT}. In: Proceedings of the IEEE/CVF
  International Conference on Computer Vision. pp. 21196--21207 (2023)

\bibitem{dudonet}
Lin, W.A., Liao, Haofu abd~Peng, C., Sun, X., Zhang, J., Luo, J., Chellappa,
  R., Kevin, Z.S.: {DuDoNet}: Dual domain network for {CT} metal artifact
  reduction. In: Proceedings of the IEEE/CVF Conference on Computer Vision and
  Pattern Recognition. pp. 10512--10521 (2019)

\bibitem{prompt-survey}
Liu, P., Yuan, W., Fu, J., Jiang, Z., Hayashi, H., Neubig, G.: Pre-train,
  prompt, and predict: {A} systematic survey of prompting methods in natural
  language processing. ACM Computing Surveys  \textbf{55}(9),  1--35 (2023)

\bibitem{swin}
Liu, Z., Lin, Y., Cao, Y., Hu, H., Wei, Y., Zhang, Z., Lin, S., Guo, B.: Swin
  transformer: Hierarchical vision transformer using shifted windows. In:
  Proceedings of the IEEE/CVF International Conference on Computer Vision. pp.
  10012--10022 (2021)

\bibitem{louis1989incomplete}
Louis, A.K., Rieder, A.: Incomplete data problems in x-ray computerized
  tomography: Ii. truncated projections and region-of-interest tomography.
  Numerische Mathematik  \textbf{56}(4),  371--383 (1989)

\bibitem{freeseed}
Ma, C., Li, Z., Zhang, Y., Zhang, J., Shan, H.: Freeseed: Frequency-band-aware
  and self-guided network for sparse-view {CT} reconstruction. In: Medical
  Image Computing and Computer Assisted Intervention -- MICCAI 2023 (2023)

\bibitem{prores}
Ma, J., Cheng, T., Wang, G., Wang, X., Zhang, Q., Zhang, L.: {P}ro{R}es:
  Exploring degradation-aware visual prompt for universal image restoration.
  arXiv preprint arXiv:2306.13653  (2023)

\bibitem{aapmmyo}
McCollough, C.H., Bartley, A.C., Carter, R.E., Chen, B., Drees, T.A., Edwards,
  P., Holmes~III, D.R., Huang, A.E., Khan, F., Leng, S., McMillan, K.L.,
  Michalak, G.J., Nunez, K.M., Yu, L., Fletcher, J.G.: Low-dose {CT} for the
  detection and classification of metastatic liver lesions: Results of the 2016
  low dose {CT} grand challenge. Medical Physics  \textbf{44}(10),  e339--e352
  (2017)

\bibitem{pytorch}
Paszke, A., Gross, S., Massa, F., Lerer, A., Bradbury, J., Chanan, G., Killeen,
  T., Lin, Z., Gimelshein, N., Antiga, L., et~al.: Py{Torch}: An imperative
  style, high-performance deep learning library. Advances in Neural Information
  Processing Systems  \textbf{32} (2019)

\bibitem{promptir}
Potlapalli, V., Zamir, S.W., Khan, S., Khan, F.S.: Prompt{IR}: Prompting for
  all-in-one blind image restoration. Advances in Neural Information Processing
  Systems  (2023)

\bibitem{stable-diffusion}
Rombach, R., Blattmann, A., Lorenz, D., Esser, P., Ommer, B.: High-resolution
  image synthesis with latent diffusion models. In: Proceedings of the IEEE/CVF
  Conference on Computer Vision and Pattern Recognition. pp. 10684--10695
  (2022)

\bibitem{torch-radon}
Ronchetti, M.: {TorchRadon}: Fast differentiable routines for computed
  tomography. arXiv preprint arXiv:2009.14788  (2020)

\bibitem{unet}
Ronneberger, O., Fischer, P., Brox, T.: U-net: Convolutional networks for
  biomedical image segmentation. In: International Conference on Medical Image
  Computing and Computer-Assisted Intervention. pp. 234--241 (2015)

\bibitem{sheng2020sequential}
Sheng, W., Zhao, X., Li, M.: A sequential regularization based image
  reconstruction method for limited-angle spectral ct. Physics in Medicine \&
  Biology  \textbf{65}(23),  235038 (2020)

\bibitem{shu2022sparse}
Shu, Z., Entezari, A.: Sparse-view and limited-angle {CT} reconstruction with
  untrained networks and deep image prior. Computer Methods and Programs in
  Biomedicine  \textbf{226},  107167 (2022)

\bibitem{dehazeformer}
Song, Y., He, Z., Qian, H., Du, X.: Vision transformers for single image
  dehazing. IEEE Transactions on Image Processing  \textbf{32},  1927--1941
  (2023)

\bibitem{dip}
Ulyanov, D., Vedaldi, A., Lempitsky, V.: Deep image prior. In: Proceedings of
  the IEEE Conference on Computer Vision and Pattern Recognition. pp.
  9446--9454 (2018)

\bibitem{transweather}
Valanarasu, J.M.J., Yasarla, R., Patel, V.M.: Transweather: Transformer-based
  restoration of images degraded by adverse weather conditions. In: Proceedings
  of the IEEE/CVF Conference on Computer Vision and Pattern Recognition. pp.
  2353--2363 (2022)

\bibitem{dudotrans}
Wang, C., Shang, K., Zhang, H., Li, Q., Zhou, S.K.: {DuDoTrans}: Dual-domain
  transformer for sparse-view {CT} reconstruction. In: Machine Learning for
  Medical Image Reconstruction. pp. 84--94 (2022)

\bibitem{wang2013meaning}
Wang, G., Yu, H.: The meaning of interior tomography. Physics in Medicine \&
  biology  \textbf{58}(16), ~R161 (2013)

\bibitem{ct-outlook}
Wang, G., Yu, H., De~Man, B.: An outlook on {X}-ray {CT} research and
  development. Medical Physics  \textbf{35}(3),  1051--1064 (2008)

\bibitem{painter}
Wang, X., Wang, W., Cao, Y., Shen, C., Huang, T.: Images speak in images: A
  generalist painter for in-context visual learning. In: Proceedings of the
  IEEE/CVF Conference on Computer Vision and Pattern Recognition. pp.
  6830--6839 (2023)

\bibitem{iqa}
Wang, Z., Bovik, A., Sheikh, H., Simoncelli, E.: Image quality assessment: from
  error visibility to structural similarity. IEEE Transactions on Image
  Processing  \textbf{13}(4),  600--612 (2004)

\bibitem{deeplesion}
Yan, K., Wang, X., Lu, L., Summers, R.M.: {DeepLesion}: Automated mining of
  large-scale lesion annotations and universal lesion detection with deep
  learning. Journal of Medical Imaging  \textbf{5}(3),  036501--036501 (2018)

\bibitem{learn++}
Zhang, Y., Chen, H., Xia, W., Chen, Y., Liu, B., Liu, Y., Sun, H., Zhou, J.:
  {LEARN++}: Recurrent dual-domain reconstruction network for compressed
  sensing {CT}. IEEE Transactions on Radiation and Plasma Medical Sciences
  \textbf{7}(2),  132--142 (2022)

\bibitem{dreamnet}
Zhang, Y., Hu, D., Hao, S., Liu, J., Quan, G., Zhang, Y., Ji, X., Chen, Y.:
  {DREAM-Net}: Deep residual error iterative minimization network for
  sparse-view {CT} reconstruction. IEEE Journal of Biomedical and Health
  Informatics  \textbf{27}(1),  480--491 (2022)

\bibitem{ddnet}
Zhang, Z., Liang, X., Dong, X., Xie, Y., Cao, G.: A sparse-view {CT}
  reconstruction method based on combination of {DenseNet} and deconvolution.
  IEEE Transactions on Medical Imaging  \textbf{37}(6),  1407--1417 (2018)

\bibitem{ms-ssim}
Zhao, H., Gallo, O., Frosio, I., Kautz, J.: Loss functions for image
  restoration with neural networks. IEEE Transactions on Computational Imaging
  \textbf{3}(1),  47--57 (2016)

\end{thebibliography}

\clearpage
\setcounter{page}{1}
\setcounter{figure}{0}
\setcounter{table}{0}
\renewcommand{\thetable}{S\arabic{table}}
\renewcommand{\thefigure}{S\arabic{figure}}

\appendix\label{sec:suppl}

\title{Supplementary Material of\\ \papertitle} 
\titlerunning{Supplementary Material of \modelname}

\author{}
\institute{}

\maketitle

\begin{abstract}
    This supplementary material includes five parts:
    (\ref{supsec:arch}) network architecture,
    (\ref{supsec:setup}) detailed experimental setup,
    (\ref{supsec:ablation}) more ablation studies,
    (\ref{supsec:visualization}) more visualization examples, 
    and (\ref{supsec:future})  future application.
\end{abstract}

\section{Detailed Network Architecture}\label{supsec:arch}
\subsection{\modelname}
Our proposed model \modelname has an hourglass architecture with skip connection, consisting of three encoding stages (including a bottleneck stage) and two decoding stages. Each stage has several transformer blocks, and detailed configurations are shown in \cref{tab:arch-proct}. As a trade-off between performance and computational burden, we do not insert W-MHSA in every transformer block; instead, we use an ``attention ratio'' parameter to indicate the percentage of transformer blocks containing W-MHSA.

Within \modelname, two separate pathways are employed, namely the source pathway and the context pathway. The source pathway is for the input source incomplete-view CT image, while the context pathway deals with the incomplete- and full-view in-context pair. Patch embedding, patch unembedding, normalization, and MLP layers are applied independently for two pathways. Specifically, to adapt the vision transformer to our image-domain CT reconstruction task, we replace the vanilla layer normalization layer with rescaled layer normalization layer~\cite{dehazeformer}, which better preserves the brightness and contrast information that is crucial for low-level vision task.

\begin{table}
\centering
\begin{minipage}[t]{0.45\linewidth}
\captionof{table}{Detailed \modelname configurations.}
\resizebox{\linewidth}{!}{
\begin{tabular}{cccccc}
    \toprule
        Configurations & Values  \\
    \midrule
    Embedding dimensions & [24,48,96,48,24] \\
    \#Transformer blocks & [8,8,8,4,4] \\
    \#Heads for W-MHSA & [2,4,6,1,1] \\
    Window size for W-MHSA & [8,8,8,8,8] \\
    MLP ratio & [2,4,4,2,2] \\
    Attention ratio & [0,1/2,1,0,0] \\
    \bottomrule
\end{tabular}}
\label{tab:arch-proct}
\end{minipage}
\hfill
\begin{minipage}[t]{0.45\linewidth}
\captionof{table}{Detailed \modelnamedd configurations.}
\resizebox{\linewidth}{!}{
\begin{tabular}{cccccc}
    \toprule
        Configurations & Values  \\
    \midrule
    Feature dimensions for $\mathcal{V}_\Phi$ & [64,128,256,512,512] \\
    Feature dimensions for $\mathcal{W}_\Psi$ & [16,32,64,128,128] \\
    \bottomrule
\end{tabular}}
\label{tab:arch-dudo}
\end{minipage}
\end{table}

\begin{figure*}[htb] 
    \centering \includegraphics[width=1.0\textwidth]{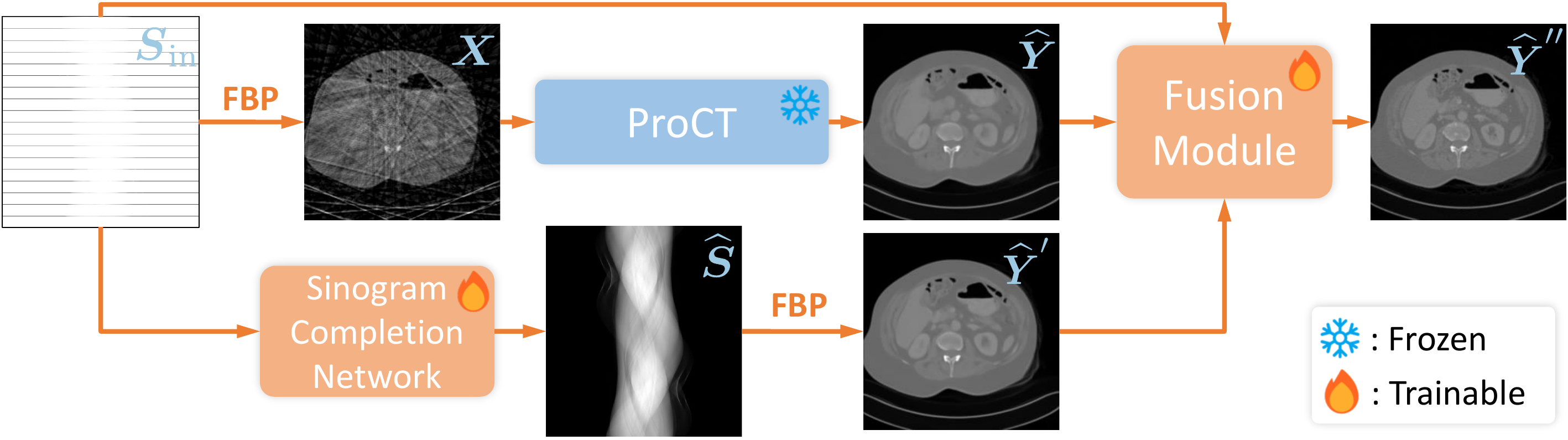}
    \caption{\modelname can be seamlessly plugged into other dual-domain frameworks as a pre-trained model providing image-domain prior knowledge. Detailed components are omitted for brevity.}
    \label{fig:model-dual-domain} 
\end{figure*}

\subsection{\modelnamedd}
To illustrate the flexibility of \modelname, we further build a simple dual-domain model termed \modelnamedd, shown in \cref{fig:model-dual-domain}, which comprises three key components: our image post-processing network \modelname ($\mathcal{U}_\Theta$), a simple U-shaped sinogram-completion convolutional network adopted from DuDoNet~\cite{dudonet} ($\mathcal{V}_\Phi$), and a light-weight fusion module ($\mathcal{W}_\Psi$). 

In a parallel configuration, $\mathcal{U}_\Theta$ processes the incomplete-view CT image $\mat{X}$ and the sampling vector $\mat{v}$ to produce an initially restored image $\widehat{\mat{Y}}$, while $\mathcal{V}_\Phi$ takes the incomplete-view sinogram $\mat{S}_\mathrm{in}$ along with the sampling mask $\mat{M}=\mathrm{diag}(\mat{v})\cdot \mat{E}$ as inputs ($\mat{E}$ is an all-ones matrix), generating a predicted full-view sinogram $\widehat{\mat{S}}$ which is subsequently transformed into an enhanced CT image $\widehat{\mat{Y}}'$. Images $\widehat{\mat{Y}}$ and $\widehat{\mat{Y}}'$, along with the input image $\mat{X}$, are fed into $\mathcal{W}_\Psi$ to give the final output $\widehat{\mat{Y}}''$. The fusion module $\mathcal{W}_\Psi$ here is also a U-Net~\cite{unet}, but with much fewer feature dimensions. These simple components already achieve promising performance, as shown in Tab.~\textcolor{red}{1} of the manuscript, yet still leave room for further improvement. More sophisticated design (\eg, prompting the sinogram completion network, enhancing the overall generalizability/transferability of \modelnamedd, and fusing the information coming from two domains, \etc) will be one of our future directions.

We apply the following dual-domain loss function to train \modelnamedd:
\begin{align}
    \mathcal{L}_\mathrm{dual}
    &= \mathcal{L}_{\ell 1}(\widehat{\mat{Y}}', \mat{Y})
    + \mathcal{L}_{\ell 1}(\widehat{\mat{Y}}'', \mat{Y})
    + \mathcal{L}_{\ell 1}(\widehat{\mat{S}}, \mat{S}).
\end{align}
Unlike the training of \modelname, we opt not to apply multi-scale structural similarity loss $\mathcal{L}_\mathrm{ssim}$ during training for our dual-domain model \modelnamedd, as we find that the learning from sinogram data inherently improves the structural similarity of the reconstructed CT images (see Tab.~\textcolor{red}{1} of the manuscript). Implementing $\mathcal{L}_\mathrm{ssim}$ would introduce unnecessary computational overhead.

\section{Detailed Experimental Setup}\label{supsec:setup}
\subsection{Data preparation}\label{supssec:data}
Two publicly available CT datasets, namely DeepLesion dataset~\cite{deeplesion} and AAPM dataset~\cite{aapmmyo}, are used in this paper. 
All the data and metadata have been anonymized to eliminate any potential personally identifiable information, and ethical approval is not required as confirmed by the attached license.

We employ the TorchRadon toolbox~\cite{torch-radon} to simulate the fan-beam CT routine under 120 kVp and 500 mA. The distance between source and rotation center, as well as detectors and rotation center, is 1075 mm each. The number of detectors is set to 672.

To simulate the photon noise in the real-world CT scenarios, we add mixed noise comprising Poisson noise at an intensity of $10^6$ and zero-mean Gaussian noise with a 0.01 standard deviation to the sinograms that generate the CT images. Note that, different from previous methods~\cite{gloredi}, the ground-truth full-view CT images for training are also generated from the noisy sinograms. This increases the difficulty of the optimization process but is more consistent with the practice of training with clinical data.

\subsection{Storage requirement of models}\label{sec:sup-storage}
We also provide a comparison of storage requirements in \cref{tab:comp-eff}. Some values are marked with ``$\times$7'', because 7 versions of the corresponding model should be stored in total to deal with all 7 settings in Tabs.~\textcolor{red}{1} and \textcolor{red}{2} of the manuscript. In contrast, only one model of \modelname needs to be stored.

\begin{table*}
    \centering
    \caption{Storage requirement for different models.}
    \resizebox{\linewidth}{!}{
    \begin{tabular}{cccccccccc}
    \toprule
    Method &DDNet &FBPConvNet &DuDoTrans &CROSS &GloReDi &FreeSeed\textsubscript{\textsc{dudo}} &UniverSeg & \modelname \\
    \midrule
    Storage (M) & 2.27$\times$7 & 131$\times$7 & 66.8$\times$7 & 139$\times$7 & 51.9$\times$7 & 184$\times$7 & 4.53 & 25.7 \\
    \bottomrule
    \end{tabular}}
    \label{tab:comp-eff}
\end{table*}

\section{More Ablation Studies}\label{supsec:ablation}
\subsection{Ablation on the view prompt}\label{supssec:ab-prompt}
Although we propose to adopt the sampling vector to form our view prompts, we find that the prompts can also be parameterized by a set of learnable vectors. However, due to the abundance of incomplete-view CT settings (\eg, SVCT can have a wide range of $N_\mathrm{view}$ choices), it is impractical to directly assign a separate learnable prompt for each setting. 

To solve this problem, for each CT scenario (SVCT and LACT), we manually divide the settings into several groups according to the reconstruction difficulty and then assign a learnable vector that ranges within $[0,1]$ for each group. For example, SVCT images with $9 \leq N_\mathrm{view} < 18$ are much harder to restore than those with $72 \leq N_\mathrm{view} < 144$. 

\begin{figure*}[!htb] 
    \centering \includegraphics[width=1.0\linewidth]{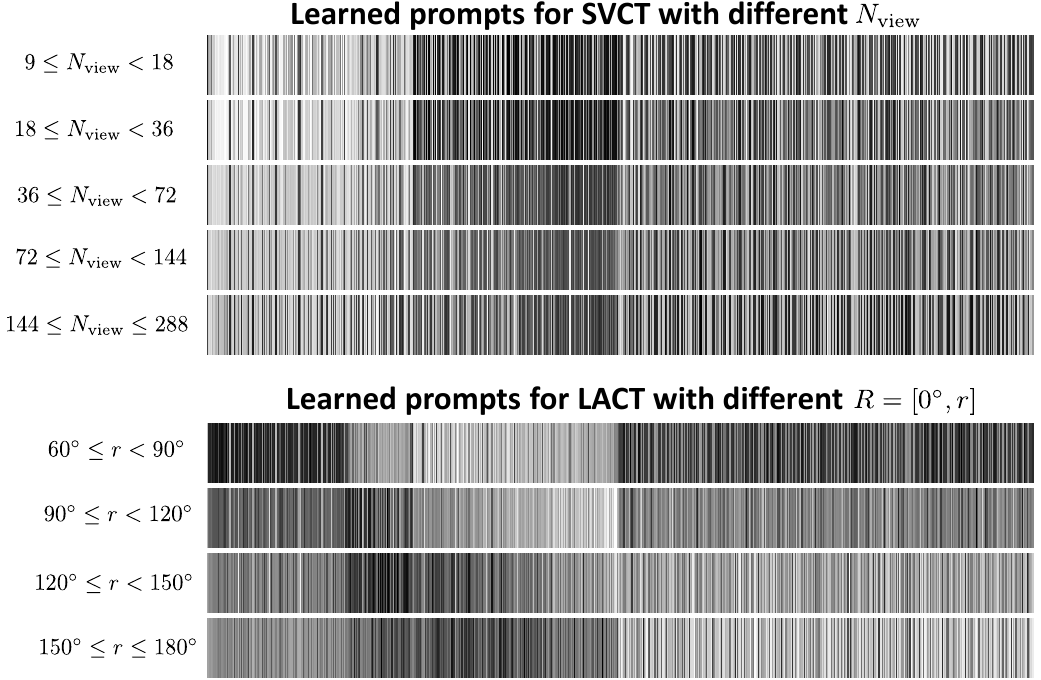}
    \caption{Visualization of learned view prompts for different incomplete-view CT settings.}
    \label{fig:prompts-visualization} 
\end{figure*}

Next, we train a variant of \modelname (dubbed ``learnable prompt'') under the same conditions as our default version. Quantitative results presented in \cref{tab:ab-learnable-prompt} indicate that, while this eliminates the need for the sampling vector, it requires manual pre-setting and leads to reduced performance compared to our default version using a deterministic prompt derived from the sampling vector. Note that, despite exhibiting certain patterns, the learned prompts still lack interpretability, as shown in \cref{fig:prompts-visualization}.
\begin{table}
    \centering
    \caption{Ablation study on the sinogram sparsity prompt.}
    \begin{tabular}{cccccc}
    \toprule
        variants
        & SVCT-PSNR & SVCT-SSIM 
        & LACT-PSNR & LACT-SSIM \\
    \midrule
    learnable prompt & 39.43 & 92.35 & 40.56 & 92.86 \\
    default & 39.73 & 93.82 & 41.33 & 94.93 \\
    \bottomrule
    \end{tabular}
    \label{tab:ab-learnable-prompt}
\end{table}

\subsection{Ablation on the loss function}\label{supssec:ab-loss}
\begin{figure}[htb] 
    \centering \includegraphics[width=0.8\linewidth]{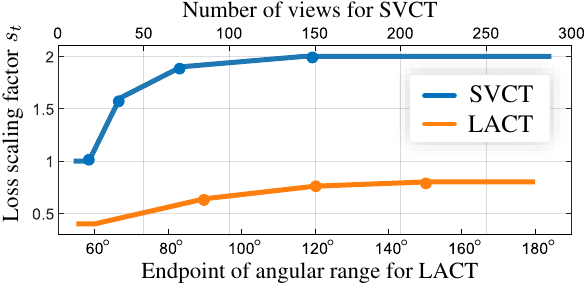}
    \caption{The loss scaling schedule $s_t$ for different incomplete-view CT settings $t$.}
    \label{fig:loss-scale} 
\end{figure}

We notice that the predicted incomplete-view CT images under difficult settings (\eg, $\nview=18,36$ in SVCT) inherently lead to larger scales of loss. To mitigate the potential bias towards these settings in the early training stage, we handcraft a scaling schedule for the loss in Eq.~(\textcolor{red}{6}), as shown in \cref{fig:loss-scale}. The scaling factor $s_t$ increases as the setting becomes easier, and its effectiveness is evident by comparing Rows 3 and 5 of \cref{tab:ab-alpha}.

\begin{table*}[!htb]
\centering
\caption{Ablation study on the balancing factor $\alpha$ and the scaling schedule $s_t$.}
\resizebox{1.0\textwidth}{!}{
\begin{tabular}{lcccccccccccccccccc} 
\toprule
{} & \multicolumn{10}{c}{\textbf{SVCT}} & \multicolumn{8}{c}{\textbf{LACT}} \\
\cmidrule(lr){2-11}  
\cmidrule(lr){12-19}
{} & \multicolumn{2}{c}{$\nview=18$} & \multicolumn{2}{c}{$\nview=36$} & \multicolumn{2}{c}{$\nview=72$} & \multicolumn{2}{c}{$\nview=144$} & \multicolumn{2}{c}{avg.} & \multicolumn{2}{c}{$R=[0^\circ,90^\circ]$} & \multicolumn{2}{c}{$R=[0^\circ,120^\circ]$} & \multicolumn{2}{c}{$R=[0^\circ,150^\circ]$} & \multicolumn{2}{c}{avg.} \\
\cmidrule(lr){2-3}
\cmidrule(lr){4-5}
\cmidrule(lr){6-7}
\cmidrule(lr){8-9}
\cmidrule(lr){10-11}  
\cmidrule(lr){12-13}
\cmidrule(lr){14-15}
\cmidrule(lr){16-17}
\cmidrule(lr){18-19}  
{\textbf{Method}} & PSNR & SSIM & PSNR & SSIM & PSNR & SSIM & PSNR & SSIM & PSNR & SSIM & PSNR & SSIM & PSNR & SSIM & PSNR & SSIM & PSNR & SSIM \\
\midrule
$\alpha=0$ w/ $s_t$
&35.36 &88.11 &38.17 &92.50 &\textbf{41.24} &\underline{96.10} &\textbf{44.22} &\textbf{97.68} &\textbf{39.75} &93.60 &\underline{37.87} &93.32 &\textbf{41.46} &95.57 &\textbf{44.73} &97.08 &\textbf{41.35} &95.32 \\
$\alpha=0.01$ w/ $s_t$
&35.35 &88.25 &38.12 &92.50 &41.19 &95.96 &\underline{44.21} &\underline{97.61} &39.72 &93.58 &37.82 &93.40 &41.38 &95.53 &44.71 &\underline{97.13} &41.30 &95.35 \\
$\alpha=0.1$ w/ $s_t$
&\underline{35.40} &88.29 &\underline{38.16} &92.72 &\underline{41.20} &\textbf{96.12} &\underline{44.21} &97.59 &\underline{39.74} &\underline{93.68} &37.84 &\underline{93.49} &\underline{41.42} &\textbf{95.79} &44.69 &\textbf{97.20} &\underline{41.32} &\textbf{95.49} \\
$\alpha=1$ w/ $s_t$
&34.37 &\underline{88.55} &37.11 &\underline{93.04} &41.10 &95.96 &43.05 &97.45 &38.90 &\textbf{93.75} &36.15 &\textbf{93.58} &40.72 &\underline{95.67} &43.96 &97.00 &40.28 &\underline{95.42} \\
$\alpha=0.1$ w/o $s_t$
&\textbf{35.96} &\textbf{89.40} &\textbf{38.74} &\textbf{93.43} &40.26 &94.91 &43.01 &96.34 &39.49 &93.52 &\textbf{38.11} &\underline{93.49} &40.68 &94.34 &42.95 &96.30 &40.58 &94.71 \\
\bottomrule
\end{tabular}
}
\label{tab:ab-alpha}
\end{table*}

In Eq.~(\textcolor{red}{5}), we introduce a factor $\alpha$ to balance the $\ell_1$ loss term $\mathcal{L}_{\ell_1}$ and the multi-scale structural similarity loss term $\mathcal{L}_\mathrm{ssim}$. We select $\alpha$ by comparing multiple versions of \modelname trained with loss scaling schedule using various $\alpha$ on the validation set, as shown in \cref{tab:ab-alpha}. While $\mathcal{L}_{\ell_1}$ encourages pixel-wise alignment between the reconstructed result and the ground-truth full-view CT image and improves PSNR, $\mathcal{L}_\mathrm{ssim}$ mainly boosts the performance in terms of statistics. We find that $\alpha=0.1$ is a good choice for balancing these two types of loss. Based on $\alpha=0.1$, we also found that \modelname trained without the proposed loss scaling schedule, dubbed ``$\alpha=0.1$ w/o $s_t$'' in \cref{tab:ab-alpha}, presents a performance drop in the average metrics, showing the effectiveness of the schedule.

\begin{figure*}[t]
    \centering
    \includegraphics[width=1\linewidth]{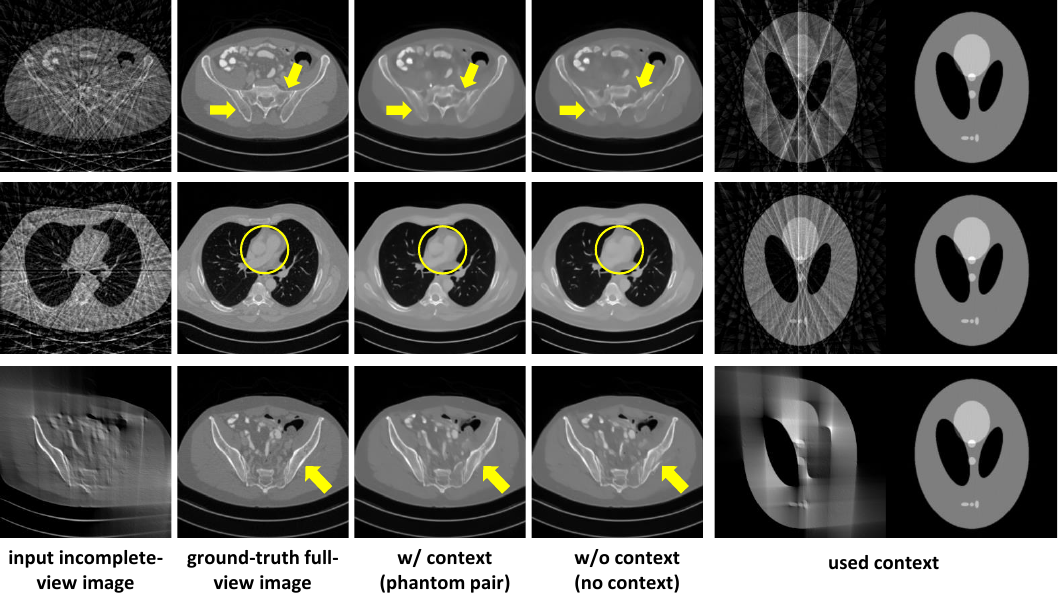}
    \caption{Contextual CT image pairs assist the restoration of some critical structures in the input incomplete-view CT image.}
    \label{fig:context-visualization} 
\end{figure*}

\subsection{Ablation on the contextual learning}\label{sec:sup-ab-context}
\paragraph{Qualitative evaluation of in-context pairs.}We first demonstrate the effectiveness of the in-context phantom pairs by providing some visualization examples of the default version of our model (``phantom pair'') and a variant that utilizes no contextual information (``no context''), as mentioned in Sec.~\textcolor{red}{4.4}. As shown in \cref{fig:context-visualization}, our model can better restore bone structure and the contour of the atrium and ventricle when using in-context pairs.

\begin{table}
    \centering
    \caption{Ablation on more configurations.}
    \resizebox{\linewidth}{!}{
    \begin{tabular}{lccccc}
    \toprule
        Configurations
        & SVCT-PSNR & SVCT-SSIM 
        & LACT-PSNR & LACT-SSIM \\
    \midrule
    (1) phantom pair w/ $\vct{p}$ (default) & 39.73 & 93.82 & 41.33 & 94.93 \\
    (2) phantom residue w/ $\vct{p}$ & 38.37 & 92.12 & 39.60 & 93.54 \\
    (3) no context w/o $\vct{p}$ & 34.16 & 84.02 & 32.94 & 84.51 \\
    \midrule
    (4) 1 real pair w/ $\vct{p}$ & \underline{39.80} & \underline{93.90} & 41.33 & 95.07 \\
    (5) 2 real pairs w/ $\vct{p}$ & \textbf{39.83} & \textbf{93.96} & \underline{41.36} & \underline{95.16} \\
    (6) 3 real pairs w/ $\vct{p}$ & \underline{39.80} & 93.76 & \textbf{41.46} & \textbf{95.23} \\
    (7) 4 real pairs & 39.20 & 93.39 & 40.92 & 94.92 \\
    \bottomrule
    \end{tabular}}
    \label{tab:ab-ncon}
\end{table}

\paragraph{Ablation on the in-context phantom pair.}Additionally, we explore two other variants as shown in Rows 2 and 3 of \cref{tab:ab-ncon}: (1) ``phantom residue w/ $\vct{p}$'', where we use the residue of the full-view phantom image and the incomplete-view phantom image as context; (2) ``no context w/o $\vct{p}$'', where neither prompt nor contextual information is given to the network. 

Although the phantom residue can also provide cues on the typical patterns of the incomplete-view CT artifact, it still leads to a performance decline compared to our default version, as can be seen in Rows 1 and 2 of \cref{tab:ab-ncon}. This is because the pixel values in the residue present a larger dynamic range and steeper landscape, leading to misalignment with the input space and adding difficulty for the convergence. Additionally, the effectiveness of the proposed view prompt and the in-context pairs can be further demonstrated by comparing Rows 1 and 3 of \cref{tab:ab-ncon}.

\paragraph{Ablation on other in-context pairs.}Despite being less consistently available, in-context pairs created from real incomplete- and full-view CT images benefit the artifact removal capability of our model, and introduce other intriguing questions that will be explored here. For example, the number of in-context pairs can affect the performance. \cref{tab:ab-ncon} shows that the best reconstruction performance is achieved when three real CT in-context pairs are given, since fewer in-context pairs may result in a biased prediction while too many pairs cause interference. 

Furthermore, we investigate the impact of using real CT in-context pairs to determine what makes effective context for \modelname. 
First, from the AAPM test set, we select three typical incomplete-view CT images visualizing various positions of the body, respectively, as the input image. For each of them, we then test our \modelname on it, trying every in-context pairs from all other images in the AAPM dataset excluding those in the test set to search for the ``best'' one that leads to the highest PSNR in the inference. 

\begin{figure}[!htb] 
    \centering \includegraphics[width=.9\linewidth]{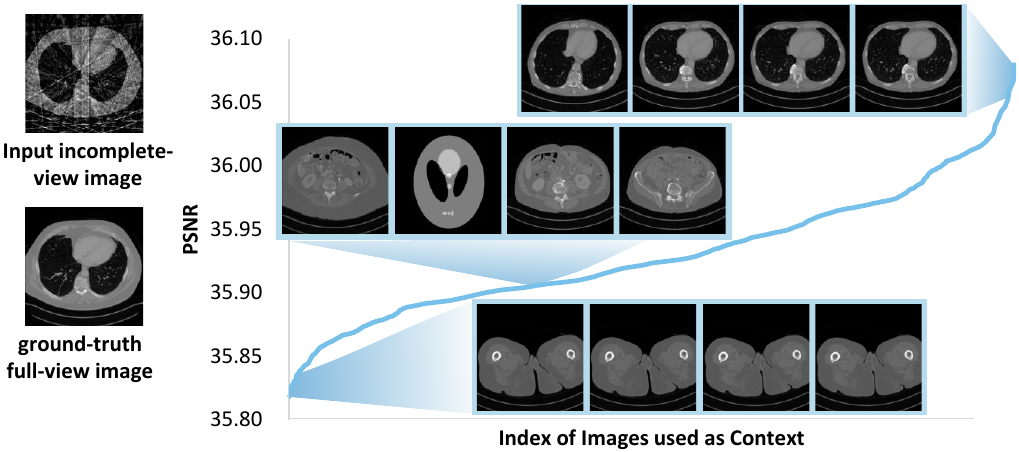}
    \caption{The more similar the content of the input image and the contextual image, the better the reconstruction of \modelname will be. Incomplete-view contextual images are omitted here for better visualization.}
    \label{fig:what-makes-context} 
\end{figure}

In \cref{fig:what-makes-context}, we visualize the contextual images corresponding to top-4 and bottom-4 PSNR values for \modelname's restoration of the input SVCT chest image ($\nview=18$). We also indicate the location of the CT phantom context and other images that lead to PSNR values similar to the CT phantom. The result suggests that contextual images that share similar content with the input image benefit \modelname for better restoration.

\begin{figure*}[h]
    \centering
    \includegraphics[width=0.9\linewidth]{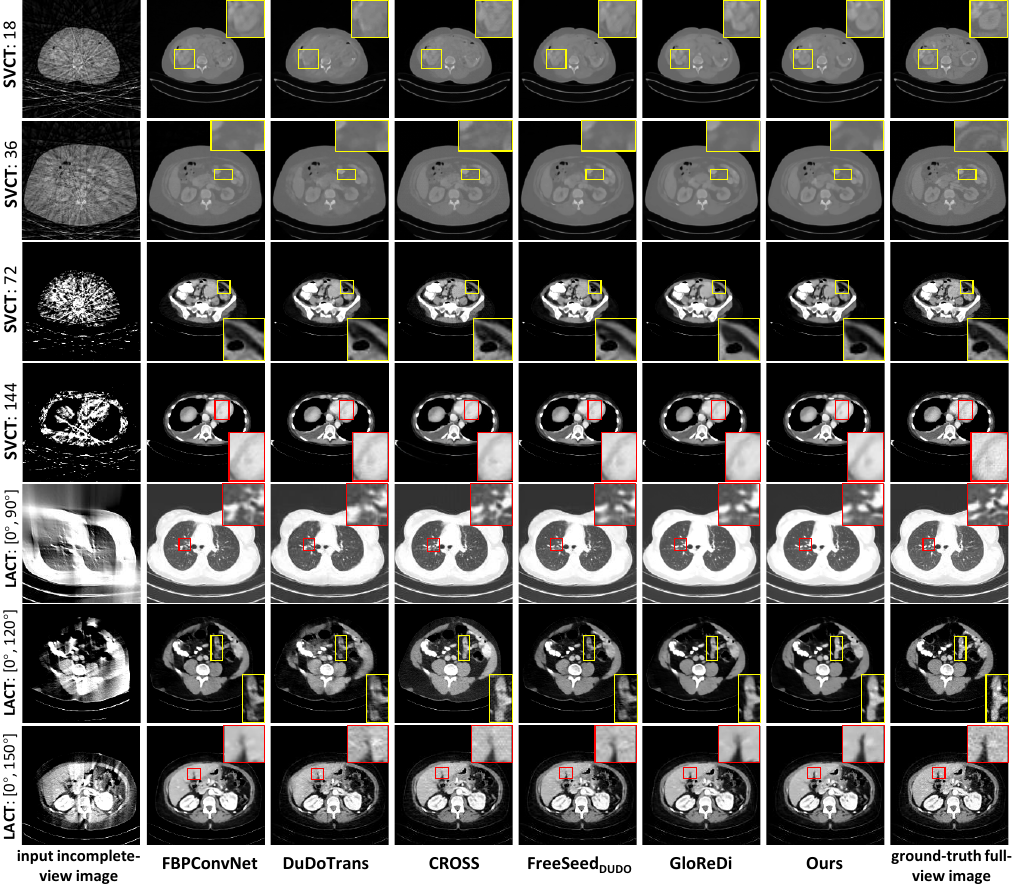}
    \caption{Visual comparison of state-of-the-art methods (from left to right) on DeepLesion dataset over different incomplete-view scenarios and settings (from top to bottom). The display windows are set to [-1000, 2000] HU for Rows 1 and 2; [-115, 185] HU for Rows 3, 4, 6 and 7; [-1250, 50] HU for Row 5. Regions of interest are zoomed in for better viewing.}
    \label{fig:comp-deeplesion-2}
    \vspace{-6pt}
\end{figure*}

\section{More Visualization Examples}\label{supsec:visualization}
\subsection{Different datasets}\label{supssec:vis-dataset}
\begin{figure*}[t]
    \centering
    \includegraphics[width=.9\linewidth]{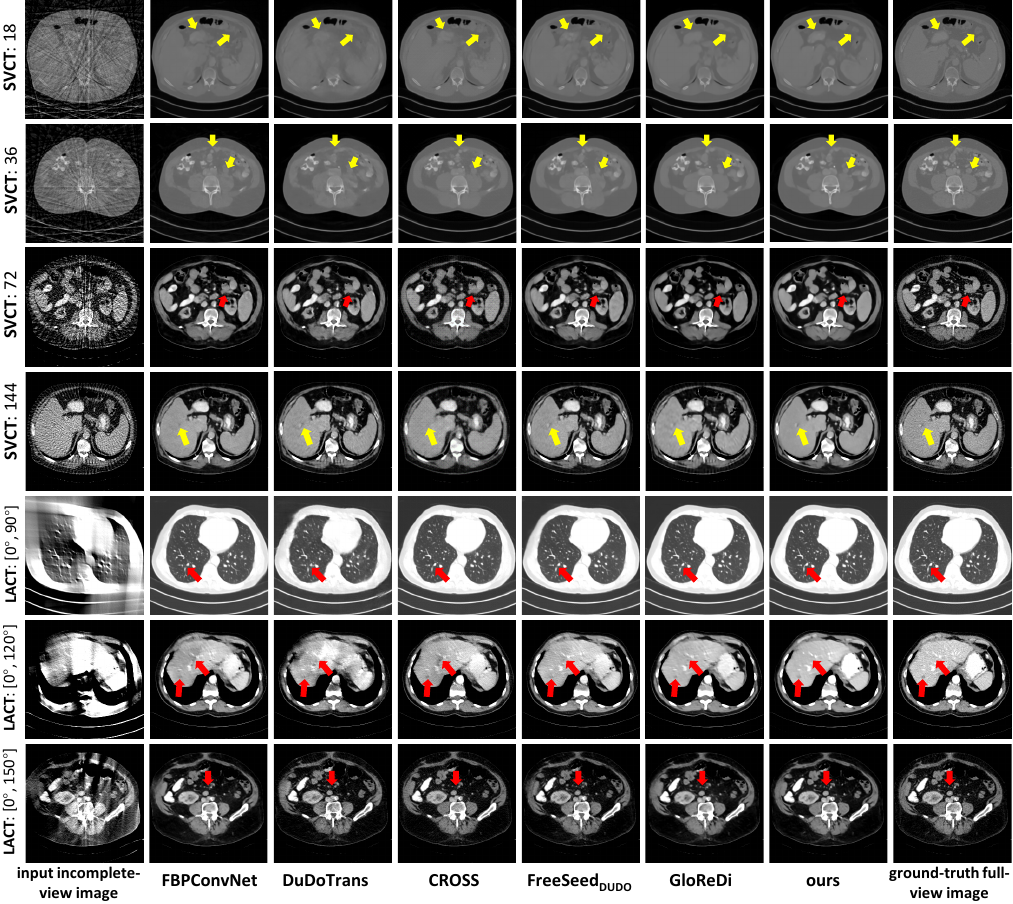}
    \caption{Visual comparison of state-of-the-art methods (from left to right) on AAPM dataset over different incomplete-view scenarios and settings (from top to bottom). The display windows are set to [-1000, 2000] HU for Rows 1 and 2; [-115, 185] HU for Rows 3, 4, 6, and 7; [-1250, 50] HU for Row 5. Regions of interest are indicated by arrows in the figures.}
    \vspace{-6pt}
    \label{fig:comp-aapm} 
\end{figure*}

We provide more visualization examples of the reconstructed incomplete-view CT images from DeepLesion dataset for state-of-the-art methods in \cref{fig:comp-deeplesion-2}, and the examples from AAPM dataset are displayed in \cref{fig:comp-aapm}. Among these methods, our proposed \modelname produces images with better quality, as can be seen in \cref{fig:comp-deeplesion-2}, where the details of the kidney (Row 1), the right upper lobe of the lung (Row 5), and other tissues (Row 6) are well maintained. In \cref{fig:comp-aapm}, some small structures missed in the images reconstructed by other methods can be restored by \modelname, as indicated by arrows in the figures.

Interestingly, dual-domain methods can capture the noise in ground-truth images. This noise is induced by the mixed noise added to the source sinogram data, resulting in the granular texture in the CT image. On the other hand, image-domain methods generally learn to remove this noise, probably working in a self-supervised manner~\cite{noise2noise}, indicating that image domain methods can potentially be robust to granular noise in the input CT image. 

\subsection{Different settings}\label{supssec:vis-settings}
\begin{figure*}[t]
    \centering
    \includegraphics[width=1\linewidth]{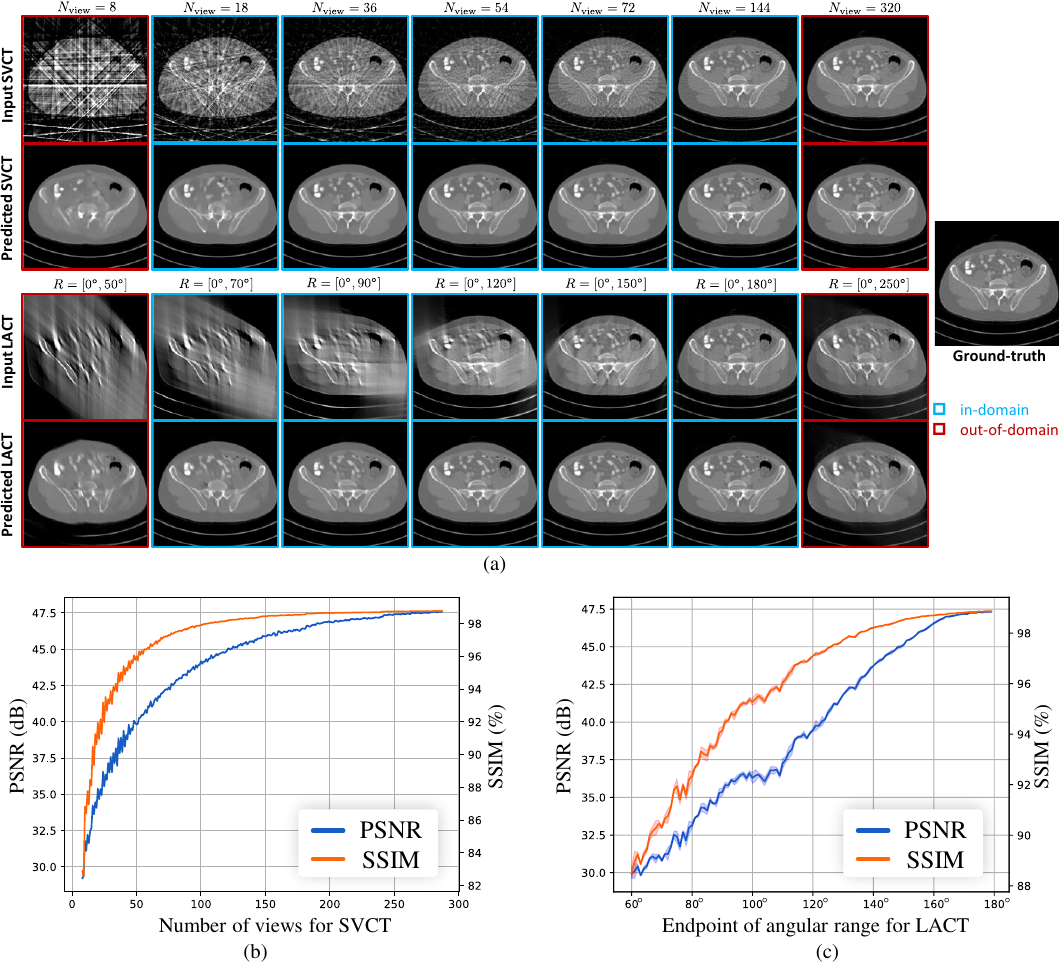}
    \caption{An example of images reconstructed by \modelname under different incomplete-view CT scenarios and settings. For each CT scenario and the corresponding exemplar image, 7 settings are visualized. (a) Reconstructed results; (b) SVCT performance profile; (c) LACT performance profile.}
    \label{fig:continuous-visualization-1} 
\end{figure*}

\begin{figure*}[!htb]
    \centering
    \includegraphics[width=1\linewidth]{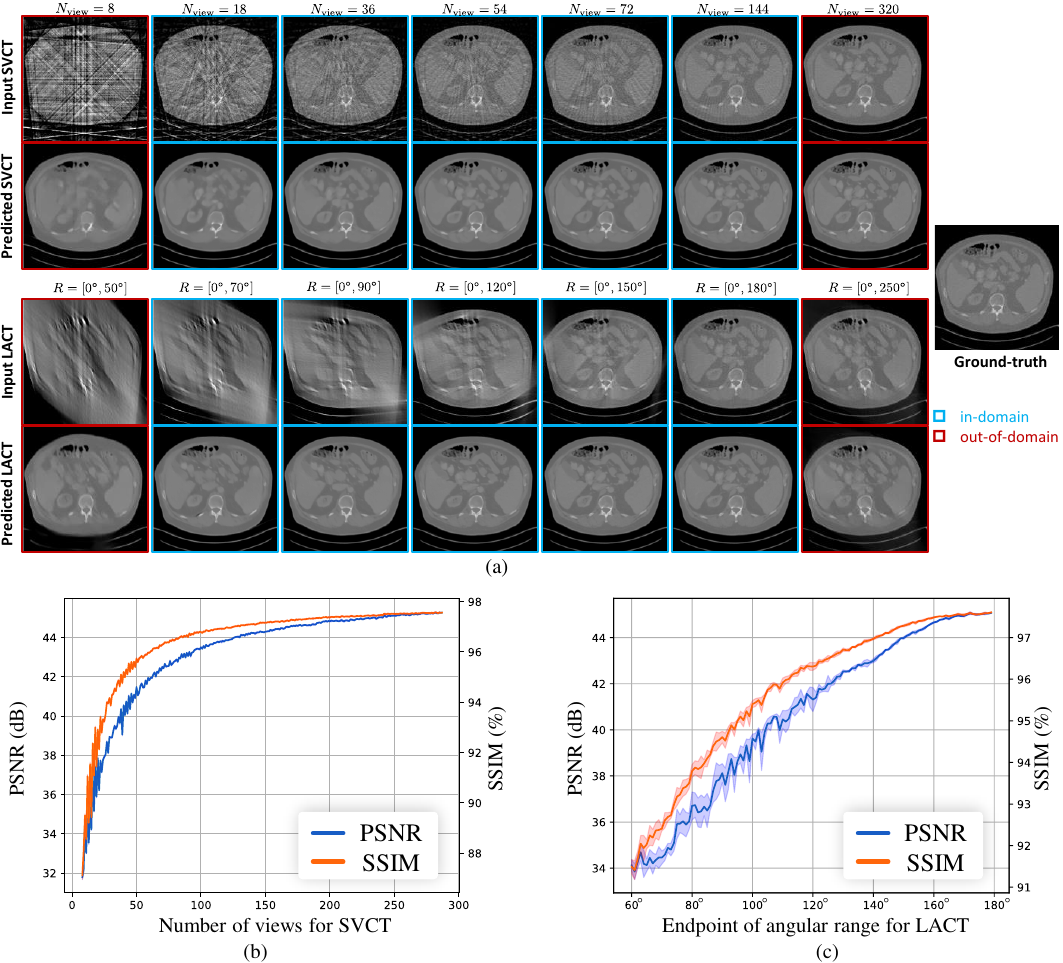}
    \caption{Another example of images reconstructed by \modelname under different incomplete-view CT scenarios and settings. For each CT scenario and the corresponding exemplar image, 7 settings are visualized. (a) Reconstructed results; (b) SVCT performance profile; (c) LACT performance profile.}
    \label{fig:continuous-visualization-2} 
\end{figure*}

 \cref{fig:continuous-visualization-1,fig:continuous-visualization-2} present the reconstructed results for two incomplete-view CT examples from AAPM dataset with different incomplete-view CT settings. Note that $R\in\{[0^\circ, 50^\circ], [0^\circ, 250^\circ]\}$ in LACT and $N_\text{view}\in\{8, 320\}$ in SVCT are out-of-domain settings. One can see that our model \modelname restores these images with satisfying visual quality in most settings, as well as the promising quantitative performance shown in the corresponding performance profiles. 

However, \modelname can fail in some cases, especially in incomplete-view CT settings with extremely few views, since most of the information is contaminated by severe artifacts. Also, in these cases, losing even one or two views can result in drastically different artifact patterns, which may be difficult for \modelname to remove.

\clearpage

\section{Future Application}\label{supsec:future}
After our multi-setting pre-training, \modelname is equipped with rich knowledge of clean CT image content and incomplete-view artifacts. This is partially validated in our experiment, where \modelname adapts to new settings with minimum fine-tuning. In line with the development in computer vision foundation models (\eg, Stable Diffusion~\cite{stable-diffusion} and its following work), we posit that \modelname also has the potential to enhance other methods by providing this prior knowledge. In addition, our \modelname can synergize with advancements in incomplete-view CT imaging to enhance imaging quality in specific scenarios. For instance, technologies like C-arms CT offer more flexible scanning solutions for intraoperative interventions and radiation therapy, where incomplete-view imaging is often encountered.

\end{document}